\documentclass[fleqn,usenatbib]{mnras}

\usepackage{newtxtext,newtxmath}
\usepackage{makecell}
\usepackage{url}
\usepackage{bm}

\usepackage[T1]{fontenc}

\DeclareRobustCommand{\VAN}[3]{#2}
\let\VANthebibliography\thebibliography
\def\thebibliography{\DeclareRobustCommand{\VAN}[3]{##3}\VANthebibliography}

\usepackage{graphicx}	
\usepackage{amsmath}	

\newcommand{\mpch}{$h^{-1}$Mpc} 
\newcommand{\kpch}{$h^{-1}$kpc} 
\newcommand{\mpchs}{$h^{-1}$Mpc$\;$} 
\newcommand{\kpchs}{$h^{-1}$kpc$\;$} 
\newcommand{\skm}{$(k/(\mathrm{s \; km}^{-1}))$}

\title[Nyx+AMBER]{The impact of varying inhomogeneous reionization histories on metrics of Ly$\alpha$ opacity}

\author[C.C. Doughty et al.]{
Caitlin C. Doughty,$^{1}$\thanks{E-mail: doughty@strw.leidenuniv.nl (CCD)}
Joseph F. Hennawi,$^{2,1}$, 
Jose O{\~n}orbe,$^{3}$,
Frederick B. Davies,$^{4}$
Zarija Luki{\'c} $^{5}$
\\
$^{1}$Leiden Observatory, Leiden University, P.O. Box 9513, 2300 RA, Leiden, the Netherlands\\
$^{2}$ Department of Physics, University of California, Santa Barbara, CA 93106, USA\\
$^{3}$ Facultad de F\'isicas, Universidad de Sevilla, Avda. Reina Mercedes s/n, Campus Reina Mercedes, E-41012 Seville, Spain \\
$^{4}$ Max-Planck-Institut f\"ur Astronomie, K\"onigstuhl 17, D-69117 Heidelberg, Germany \\
$^{5}$ Lawrence Berkeley National Laboratory, CA 94720-8139, USA 
}

\date{Accepted XXX. Received YYY; in original form ZZZ}

\pubyear{2023}

\begin{document}
\label{firstpage}
\pagerange{\pageref{firstpage}--\pageref{lastpage}}
\maketitle

\begin{abstract} The epoch of hydrogen reionization is complete by $z=5$, but its progression at higher redshifts is uncertain. Measurements of Ly$\alpha$ forest opacity show large scatter at $z<6$, suggestive of spatial fluctuations in neutral fraction ($x_\mathrm{HI}$), temperature, or ionizing background, either individually or in combination. However, these effects are degenerate, necessitating modeling these physics in tandem in order to properly interpret the observations. We begin this process by developing a framework for modeling the reionization history and associated temperature fluctuations, with the intention of incorporating ionizing background fluctuations at a later time. To do this, we generate several reionization histories using semi-numerical code AMBER, selecting histories with volume-weighted neutral fractions that adhere to the observed CMB optical depth and dark pixel fractions. Implementing these histories in the \texttt{Nyx} cosmological hydrodynamics code, we examine the evolution of gas within the simulation, and the associated metrics of the Ly$\alpha$ forest opacity. We find that the pressure smoothing scale within the IGM is strongly correlated with the adiabatic index of the temperature-density relation. We find that while models with 20,000 K photoheating at reionization are better able to reproduce the shape of the observed $z=5$ 1D flux power spectrum than those with 10,000 K, they fail to match the highest wavenumbers. The simulated autocorrelation function and optical depth distributions are systematically low and narrow, respectively, compared to the observed values, but are in better agreement when the reionization history is longer in duration, more symmetric in its distribution of reionization redshifts, or if there are remaining neutral regions at $z<6$. The systematically low variance likely requires the addition of a fluctuating UVB.
\end{abstract}




\section{Introduction}\label{sec:introduction}
The epoch of reionization (EoR) is the last major phase transition in the Universe's history, when the neutral intergalactic medium (IGM) was ionized for the first time since before recombination~\citep[e.g.][]{loeb01}. Initiating when the first stars began to form, it could not progress in earnest until a suitable star formation rate density was achieved, since the production rate of ionizing photons had to outstrip the rate of recombination.

While the amount of neutral material could be measured through observations of hydrogen Lyman series absorption, the high mean density of the IGM at $z>3$ combined with the high oscillator strength of the Ly$\alpha$ transition at $z=5$ means saturation occurs even for sub-percent neutral fractions, preventing direct determination of its timeline. The best constraints on the EoR timeline have come from measurements of the Cosmic Microwave Background (CMB) from the Planck experiment, giving $\tau_\mathrm{CMB}=0.054\pm0.007$~\citep{planck20}. This was combined with assumed reionization histories using the FlexKnot model~\citep{millea18}, which treats the timeline as a series of ``knots'' in $z$ and $x_\mathrm{HI}$ space, and used to establish an approximate timeline with a midpoint at $z=7.7\pm0.7$. The next best constraints will eventually come from 21 cm experiments like the Hydrogen Epoch of Reionization Array~\citep{deboer17}, which will be able to tomographically map neutral hydrogen throughout the reionization epoch, allowing for a thorough assessment of the history. However, while some experiments have preliminary results placing constraints on pre-reionization X-ray heating~\citep{hera23}, tomographic results are still a ways off, with much work currently going towards diminishing significant foreground contamination from interlying celestial structures as well as Earth-based signals.

Currently, there is a growing body of high-$z$ quasar and galaxy observations that are also being used to chart the progression of reionization. QSO damping wings at $ 7< z < 7.5$ are producing a wide range of neutral fractions, from almost 0 to 0.9 including the uncertainties, depending on the quasar and analysis method~\citep{greig17,banados18,davies18,greig19,wang20,yang20,greig22}. The large variation in these measurements, especially towards very low neutral fractions, may indicate issues with the inference procedure, or that the quasars are located in biased regions that are more ionized at early times~\citep[e.g.][]{costa14}. Using galaxies in a similar way may hold promise, though there is additional uncertainty with determining their intrinsic spectra, which are more variable than those of quasars~\citep{keating23}. Additionally, IGM damping wings observed in galaxy spectra may be contaminated by gas in the background galaxy's intergalactic or circumgalactic media~\citep{heintz23}.

For the last stages of reionization occurring at $z<6$, it is possible to measure a nonzero Ly$\alpha$ flux $F$, and learn about the neutral fraction by considering the amount of ``dark'' $F\sim0$ regions compared to those with $F>0$. The most certain results are from~\citet{mcgreer15}, which by quantification of the dark pixel fraction established upper limits on the volume averaged abundance of neutral hydrogen to 0.09 at $z=5.6$. Other studies have achieved model-dependent results through evaluation of Ly$\alpha$ and $\beta$ dark gaps in tandem with simulations~\citep{zhu22, jin23}.

The full distribution of Ly$\alpha$ opacities contains more information than dark pixels alone, the values being impacted by the field values of neutral fraction, gas density, temperature, and the ultraviolet background (UVB). For a fully ionized IGM, the Ly$\alpha$ opacity ($-\ln{F}$) will vary with these values according to
\begin{equation}
\tau \propto n_\mathrm{HI} \propto T^{-0.7} \propto \Gamma_\mathrm{HI}^{-1}
\end{equation}
These measurements have been used to constrain the neutral fraction, mean free path, and photoionization rate~\citep{bosman22,gaikwad23,zhu23}. The fluctuations in these fields also contribute to scale-dependent variations in power, glimpsed through clustering measures like the Ly$\alpha$ power spectrum and flux autocorrelation function~\citep{viel13b,walther18,boera19,karacayli22}.

In order to properly interpret observations, it is essential to compare to predictions, and for such a complex process as reionization, this necessitates the use of numerical simulations. These simulations must be able to accurately capture \emph{(1)} the structure at all relevant scales down to the pressure smoothing scale that characterizes the small-scale cutoff in power due to the IGM thermal history~\citep{kulkarni15,puchwein23,doughty23} and \emph{(2)} the inhomogeneous reionization and post-reionization thermal state (including the temperature at mean density and the slope of the temperature density relation). However, given the known degeneracies between reionization timing, the amount of photoheating to expect, and the UVB amplitude compounded with the overall uncertainty regarding the reionization timeline, it is important to consider a wider range of reionization histories, simulated using an appropriate code, and at an appropriate resolution to capture the pressure smoothing.

Here, we begin this work by using the semi-numerical code AMBER to create a variety of inhomogeneous reionization histories that are compatible with the best available constraints on reionization: the \emph{Planck} CMB results and model-independent dark pixel fractions. Using a small set of sample histories, and varying the heat injection within the cosmological hydrodynamics code \texttt{Nyx}, we demonstrate how the parameters of the reionization histories impact gas within the IGM and lead to degenerate conclusions results for a preferred ionization history when examined using a simplistic ``$\chi$-by-eye'' evaluation technique. In Section~\ref{sec:simulations} we begin by describing our methods, including the hydrodynamical simulation setup, modeling of inhomogeneous reionization, and justification for our chosen test $x_\mathrm{HI,v}(z)$ histories. In Section~\ref{sec:results} we first explore how the histories affect the temperature and density evolution of gas in the IGM, including the pressure smoothing. We then delve into the effect on the observable Ly$\alpha$ opacities and compare them to a subset of existing observations. We conclude the work by synthesizing the results in Section~\ref{sec:discussion} and presenting a summary in Section~\ref{sec:conclusions}.

\section{Simulations}\label{sec:simulations}

In order to model reionization, we must include separate components to evolve the formation of structure as well as the actual ionization process. The reionization of neutral hydrogen is a prolonged process, with the bulk  occurring below $z=9$. It is induced in overdense regions earlier due to the greater abundance of ionizing photons there, resulting in a patchy field of H II. This generates fluctuations in temperature as well, both impacting the opacity in the Ly$\alpha$ forest. In this section we describe the method for modeling this patchy reionization, and the details of the histories we choose.

\subsection{Modeling an inhomogeneous reionization}\label{sssec:amber}

Reionization can be modeled through the inclusion of \emph{in-situ} radiative transfer~\citep[e.g.][]{rosdahl18,rosdahl22,kannan22}, or by post-processing of the volume~\citep[e.g.][]{puchwein23}. More cheaply, semi-analytical methods may be used to generate realistic ionization fields, as in~\citet{battaglia13}, or tools like {\sc 21cmfast}~\citep{mesinger11} and FASTPM~\citep{feng16}.

To generate the reionization field for our simulations we elect to use the abundance-matching model AMBER~\citep{trac22}, which creates a gridded reionization field by matching a selected mass-weighted reionization history to a rank-ordered pseudo-radiation field. This matching is based on the assumption that grid cells encountering a stronger radiation field will reionize first. The first step is determining a reionization history, which is parametrized in terms of a midpoint $z_\mathrm{mid}$, duration $\Delta z$, and asymmetry $A_z$. The midpoint is the redshift where the mass-weighted ionized fraction $x_\mathrm{HII,m}$ equals 0.5. ``Early'' and ``late'' points in the history are defined as $x_\mathrm{HII,m}=0.05$ and $x_\mathrm{HII,m}=0.95$, respectively, and calculated as
\begin{equation}
    z_\mathrm{early} = z_\mathrm{mid} + \frac{\Delta z A_z}{1+A_z} \: \: ; \: \: z_\mathrm{late} = z_\mathrm{mid} - \frac{\Delta z A_z}{1+A_z}
\end{equation}
Using these definitions, the duration is equal to the difference between $z_\mathrm{early}$ and $z_\mathrm{late}$, and the asymmetry is the ratio $(z_\mathrm{early}-z_\mathrm{mid})/(z_\mathrm{mid}-z_\mathrm{late})$. A Weibull function~\citep{weibull51} is used to interpolate between the three fixed points $z_\mathrm{mid}$, $z_\mathrm{early}$, and $z_\mathrm{late}$ and arrive at a fully defined history. 

\begin{figure}
\includegraphics[width=0.5\textwidth]{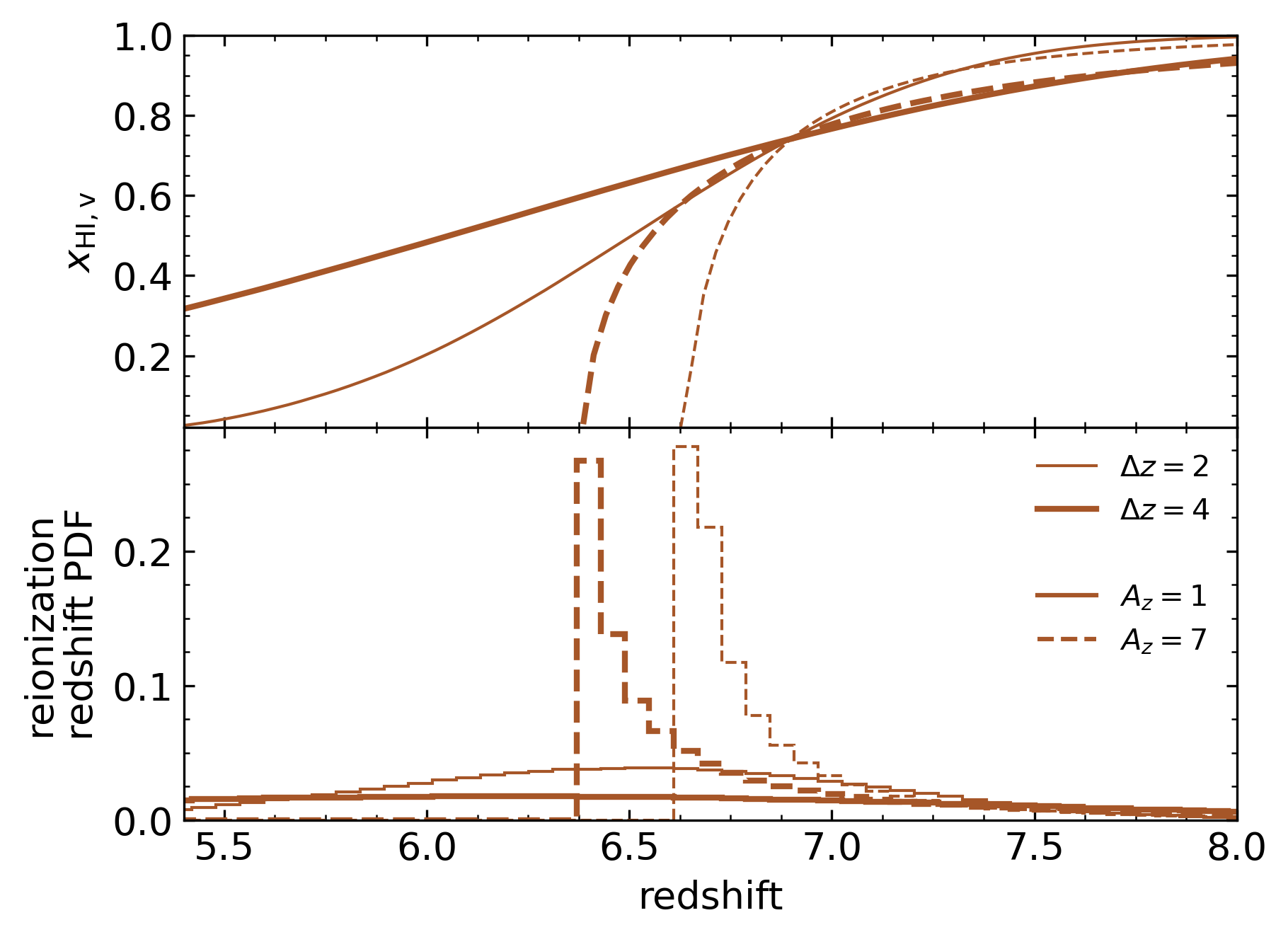}
\caption{The volume-weighted neutral fraction (\emph{top}) and PDFs of the reionization redshift (\emph{bottom}) for four $z_\mathrm{mid}=6.9$ example reionization histories from AMBER, demonstrating the impact of varying the duration ($\Delta z$) and asymmetry ($A_z$) parameters. With all other parameters held constant, the (1) duration dictates the peak fraction being ionized, and (2) asymmetry dictates whether the distribution of $z_\mathrm{re}$ is symmetric about the peak. In practice, an asymmetric model leads to a ``tail'' in $x_\mathrm{HI,v}$ out to lower $z$. While not shown here, the mass-weighted midpoint shifts the ($x_\mathrm{HI,v}$,$z$) intersection point horizontally.}
\label{fig:amber_examples}
\end{figure}

In the base AMBER code, the density field is acquired by first evolving a Gaussian random field (adjusted to match the cosmology-determined power spectrum) using linear perturbation theory~\citep[LPT;][]{bouchet95,scoccimarro98}, with the option to use either the Zel'dovich approximation~\citep{zeldovich70} or the more accurate second order LPT (2LPT). Assignment of the particles to a grid is performed using two staggered, interlaced meshes to improve the accuracy of the overdensity fields, and avoid effects such as aliasing~\citep{hockney88}. We do not use the built-in functionality of AMBER to generate initial conditions from a Gaussian random field. Instead we extract the $z=0$ linear density field from our initial conditions generator (see Section~\ref{ssec:nyx} for more details) and use AMBER to evolve it using 2LPT to $z=z_\mathrm{mid}$.

From the overdensity field, the excursion set formalism is used to model the collapsed mass fraction within the simulation volume, and extended Press-Schechter theory is used to derive a halo mass function and the resultant halo density field. To extend this to the radiation field, the background is defined as usual
\begin{equation}
    \Gamma_\mathrm{HI}(\textbf{x)} = 4 \pi \int^\infty_{\nu_\mathrm{HI}} \frac{J_\nu (\nu, \textbf{x})}{h \nu} \sigma_\mathrm{HI}(\nu) \mathrm{d}\nu 
\end{equation}
where $\sigma_\mathrm{HI}$ is the cross-section for hydrogen photoionization. $J_\mathrm{\nu}$ is the specific intensity, itself given by
\begin{equation}
    J_\mathrm{\nu}(\nu, \textbf{x}) = \int \frac{S_\nu(\nu, \textbf{x'})}{4 \pi |\textbf{x}-\textbf{x}'|^2} \exp \left( -\frac{|\textbf{x}-\textbf{x}'|}{\lambda_\mathrm{mfp}} \right) \mathrm{d^3x'}
\end{equation}
with source function $S_\nu$, and the mean free path of ionizing photons $\lambda_\mathrm{mfp}$. Recent studies of the mean free path suggest significant evolution between $5 < z < 6$~\citep{becker21,gaikwad23,zhu23}, but for the majority of the reionization process that occurs at $z>6$, this value is uncertain. The mean free path $\lambda_\mathrm{mfp}$ is defined using the standard theoretical definition, the distance over which the transmitted flux fraction drops to $1/e$. Given that the mean free path mainly affects the topology of reionization rather than the history, we simply use a constant 3 \mpchs for all our histories, and leave a more detailed consideration for future work.

The source function,
\begin{equation}\label{eq:source_function}
    S_\nu (\nu, \textbf{x}) = f_\mathrm{esc} \epsilon_\nu (\nu) \rho_\mathrm{SFR}(\textbf{x}),
\end{equation}
is dependent on several astrophysical variables such as the radiation escape fraction $f_\mathrm{esc}$, the radiative energy per unit star formation rate (SFR) per $\nu$. The star formation rate density is in turn given by
\begin{equation}\label{eq:sfr_density}
    \rho_\mathrm{SFR} (\textbf{x}) = \frac{f_\mathrm{star} \rho_\mathrm{halo}(\textbf{x})}{\tau_\mathrm{star}}
\end{equation}
where $\rho_\mathrm{halo}$ is the halo mass density field, and $f_\mathrm{star}$ and $\tau_\mathrm{star}$ are the star formation efficiency and timescale, respectively. Equations~\ref{eq:source_function} and~\ref{eq:sfr_density} are not themselves a part of the reionization parametrization, and the variables therein are simply tuned to produce a history with the desired values. 

All $N$ cells are assigned a rank order $n$ based on the values of the radiation field, and the mass fraction is calculated for each rank
\begin{equation}
\frac{\sum^n_{m=1} 1 + \delta_m(z)}{\sum^N_{m=1} 1 + \delta_m(z)} = \bar{x}_i (z)
\end{equation}
where $\bar{x}_i (z)$ is the ionized fraction at redshift $z$ and $\delta_m$ is the overdensity of the $m$th cell at $z$. For each mass fraction, the reionization redshift can be determined by passing the fraction $\bar {x_i}$ to the inverted Weibull function. This uses the density field at $z_\mathrm{mid}$, but a few iterative steps are performed to adjust these values to the field at $z=z_\mathrm{re}$ assuming linear growth of the overdensities from the reionization midpoint.

We show some example histories in the top panel of Figure~\ref{fig:amber_examples}, for a constant reionization midpoint of $z=6.9$ and two choices each for asymmetry and duration. The midpoint plays the primary role in positioning the reionization history, here producing a volume-weighted neutral fraction of approximately 75 per cent at $z=6.9$. For a constant asymmetry parameter of $A_\mathrm{z}=7$ (dashed lines), a longer duration of $\Delta z=4$ compared to 2 increases the time between the starting and ending redshifts, so reduces the steepness of the history. For a constant duration $\Delta z=2$ (thin lines), a greater asymmetry steepens the history profile and in particular removes any lingering neutral regions at late times.

To visualize this another way, we show the normalized PDFs of the redshifts at which the cells reionize in these example models (bottom panel Figure~\ref{fig:amber_examples}). Here, the distributions for the two symmetric models ($A_\mathrm{z}=1$) peak at the center of an approximately normal distribution, with a roughly equal numbers of cells reionizing above and below the peak. Their peaks also contain only a small fraction of the total cells, $<10$ percent. The asymmetric distributions on the other hand have many more occurring at $z>z_\mathrm{peak}$ and $\sim30$ percent in the peak bin itself. The peaks do not occur at the mass-weighted midpoint, but rather at a lower redshift. It is apparent that the models with asymmetric histories will experience an abrupt ionizing event, coupled with a heat injection that may result in a relatively uniform temperature field and a more coherent pressure smoothing scale.

\begin{table*}
  \centering
  \caption{A summary of the simulations used in Section~\ref{sec:results}. The $z_\mathrm{re}$ fields have a coarse grid of $128^3$ and are generated using AMBER to abundance match the radiation field to an arbitrary reionization history, as described in Section~\ref{sssec:amber}. $z_\mathrm{early}$, $z_\mathrm{mid}$, and $z_\mathrm{late}$ are the points that define the arbitrary history in terms of mass-weighted H I fractions. All of the runs use $L_\mathrm{box}=20$ \mpchs with a fixed grid resolution $2048^3$ cells, corresponding to spatial resolution $\Delta x=10$ \kpch. The naming convention distinguishes between midpoints (Late $z\sim6$, Middle $z\sim7$, and Early $z\sim8$, durations (Short $\Delta z\leq3$ or Long $\Delta z > 3$), and asymmetry (Asymmetric $A_z>1$ or Symmetric $A_z=1$). The heat injection is further indicated using Cold (10000 K) or Hot (20000 K).} \label{tab:reionization_model_details}
  \begin{tabular}{ l c c c c c c c c}
    \hline
     name & \makecell{$z_\mathrm{early}$ \\ ($x_\mathrm{HI,m}=0.95$)} & \makecell{$z_\mathrm{mid}$ \\ ($x_\mathrm{HI,m}=0.5$)} & 
     \makecell{$z_\mathrm{late}$ \\ ($x_\mathrm{HI,m}=0.05$)} & 
     \makecell{$z_\mathrm{end}$ \\ ($x_\mathrm{HI,m}=0.0$)} & 
     $\Delta z$ & $A_\mathrm{z}$ &
     \makecell{$\Delta T_\mathrm{re}$ \\ (K)} &  $\tau_\mathrm{e}$ \\
     \hline
      \textsl{LateShortAsym} & 7.9 & 6.1 & 5.9 & 5.9 & 2 & 8 & 1e4, 2e4 & 0.0398 \\
      \textsl{MidShortAsym} & 8.7 & 6.9 & 6.7 & 6.7 & 2 & 8 & 1e4, 2e4 & 0.0475 \\
      \textsl{MidLongAsym} & 21.2 & 7.3 & 6.2 & 6.2 & 15 & 13 & 1e4, 2e4 & 0.0596 \\
      \textsl{EarlyLongAsym} & 21.8 & 7.9 & 6.8 & 6.8 & 15 & 13 & 1e4, 2e4 & 0.0661 \\
      \textsl{EarlyShortSym} & 9.7 & 8.2 & 6.7 & 5.4 & 3 & 1 & 1e4, 2e4 & 0.0549 \\
      \textsl{EarlyLongSym} & 10.2 & 8.2 & 6.2 & 4.4 & 4 & 1 & 1e4, 2e4 & 0.0529 \\
    \hline
  \end{tabular}
\end{table*}

\begin{figure}
\hspace{-0.5cm}
\includegraphics[width=0.5\textwidth]{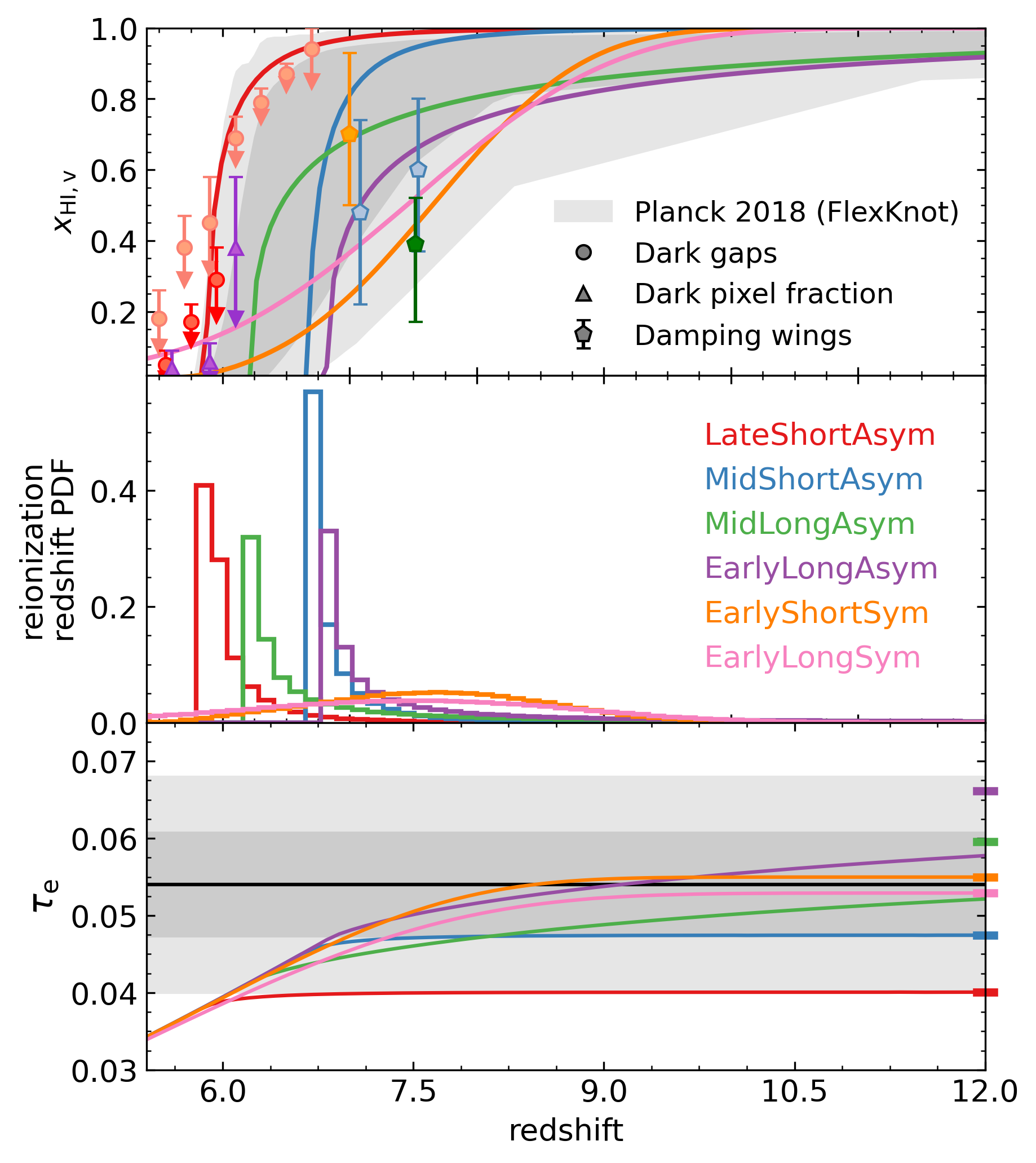}
\caption{\emph{Top:} Reionization histories for our selected models generated using AMBER. The dark and light gray shaded regions indicate the constraints from~\citet{planck20} using the FlexKnot model. Overplotted are observationally-inferred values from~\citet{mcgreer15} (purple),~\citet{zhu22} (red),~\citet{jin23} (salmon), ~\citet{davies18} (blue),~\citet{wang20} (orange), and ~\citet{yang20} (green).  \emph{Middle:} PDFs of the reionization redshifts for each model. Visually, the most major differences are between symmetric \textsl{Sym} and asymmetric \textsl{Asym} models, with the latter having high peaks. \emph{Bottom:} The Thompson scattering optical depths for the reionization models in Figure~\ref{fig:xHI_plot}, calculated according to Equation~\ref{eq:tau_e}. The dark and light gray shaded regions indicate the $1$ and $2\sigma$ limits from~\citet{planck20}, and the colored ticks on the right y-axis show the full predicted $\tau_e$ when integrated out to the surface of last scattering. Broadly, a model that is more reionized earlier has a higher $\tau_e$, since there are more free electrons in the IGM to contribute to scattering.}
\label{fig:xHI_plot}
\end{figure}

\subsection{Selecting reionization histories}
Figure~\ref{fig:xHI_plot} shows our selected reionization histories as the volume-weighted H I neutral fraction ($x_\mathrm{HI,v}$), overplotted with select observational constraints,\footnote{We note again that the model parameters describe the \emph{mass-weighted} reionization history, while the history in Figure~\ref{fig:xHI_plot} is volume-weighted.} including~\citet{planck20} 2$\sigma$ limits indicated in gray. Given our goal of probing the parameter space suggested by the strongest constraints~\citep[taken here to be \emph{Planck} FlexKnot and][]{mcgreer15}, we select several combinations of ($z_\mathrm{mid}$, $\Delta z$, $A_z$) to span most of this swath; the details of these models are presented in Table~\ref{tab:reionization_model_details}. For the naming convention, we use \textsl{Early}, \textsl{Mid}, and \textsl{Late} to indicate early ($z\sim8$), middling ($z\sim7$), and late ($z\sim6$) mass-weighted midpoints. For duration, we use \textsl{Short} ($\Delta z<4$) or \textsl{Long} ($\Delta z\geq4$), a somewhat arbitrary designation. For asymmetry we use \textsl{Asym} ($A_z>1$) and \textsl{Sym} ($A_z=1$). Lastly, we use hot ($\Delta T_\mathrm{re}=20,000$ K) and cold ($\Delta T_\mathrm{re}=10,000$ K) to refer to the amount of heat injection as a region reionizes (see Section~\ref{ssec:nyx} for details of this implementation).

We examine values within a reasonable parameter space $6.1<z_\mathrm{mid}<8.2$, $2<\Delta z<15$, and $1<A_\mathrm{z}<13$. Within this space, there are many possible histories, but we manually select six that extend from extremely late, short histories (upper left of panel \emph{a}) to earlier, and more symmetric ones (lower right of panel\emph{a}). Most of our selected histories do not violate the $2\sigma$ limits of FlexKnot with the exception of the \textsl{EarlyLongSym} model, which approximately matches the~\citet{mcgreer15} limits. However, the true endpoints $z_\mathrm{end}$ where the volume-weighted neutral fraction reaches zero sometimes occur quite late, $z_\mathrm{end}=5.4$ and 4.4 for \textsl{EarlyShortSym} and EarlyLongSym, respectively. \textsl{EarlyLongSym} in particular is quite late to reionize fully, but we retain it as an extreme example history.

Besides~\citet{mcgreer15} in purple, the other colored points in Figure~\ref{fig:xHI_plot} indicate inferences from Ly$\alpha$ and Ly$\beta$ dark gaps~\citep{zhu22,jin23}. For $z\geq7$, we show select results from analyses of quasar damping wings~\citep{davies18,banados18,greig19}. Other methods of inferring this history, whose results we do not include here, include the Ly$\alpha$ equivalent width distribution~\citep[e.g.][]{mason18,hoag19,bruton23} and Ly$\alpha$ clustering~\citep[e.g.][]{ouchi10}. Many of these $z>7$ observationally-inferred $x_\mathrm{HI,v}$ values are in some contention with one another, related to systematic discrepancies between the methods but also potentially due to the patchy topology of reionization and/or the biased nature of certain environments (e.g. around quasars). However, these other results are broadly in support of the \emph{Planck} limits, and the histories we have chosen.

In the center panel of Figure~\ref{fig:xHI_plot} we show the normalized PDFs of the redshifts at which the cells reionized. The four asymmetric models (\textsl{Asym}) are highly peaked, so it is expected that these models will resemble a ``flash'' or instantaneous reionization history, where a large amount of heat is injected to the IGM at a single redshift. The symmetric models (\textsl{Sym}) show a ``normal'' distribution of $z_\mathrm{re}$, which will lead to a gradual injection of heat in the simulations and a wide distribution of pressure smoothing lengths in the IGM.

Using the reionization fields generated in AMBER, we calculate the optical depth to Thompson scattering of the cosmic microwave background, $\tau_e$. To accomplish this, we use
\begin{equation}\label{eq:tau_e}
    \tau_\mathrm{e} = \int_{0}^{z_\mathrm{ls}} \frac{c\;\sigma_\mathrm{T} \; n_e\left(z\right)}{\left(1+z\right) \; H\left(z\right)} \mathrm{d}z 
\end{equation}
where $z_\mathrm{ls}$ is the redshift of last scattering ($z\sim1100$), $\sigma_T$ is the Thompson scattering cross-section, $n_e$ is the electron density, and $H(z)$ is the Hubble parameter at redshift $z$. For the electron density, we assume contributions from both H II and He II for $z>3$, and add contributions from He III for $z\leq3$, assuming that quasars have reionized all the He II by this point in time.

With this approximation we arrive at the curves in the bottom panel of Figure~\ref{fig:xHI_plot}. The lowest and highest $\tau_\mathrm{e}$ correspond to models \textsl{LateShortAsym} and \textsl{EarlyLongAsym}, whose largest differences are in their midpoints and durations. \textsl{LateShortAsym} initiates later and is much shorter, leading to a smaller integrated electron density along the line of sight compared to the \textsl{EarlyLongAsym} model, which initiates quite early. The two models closest to the \emph{Planck} result are \textsl{EarlyShortSym} and \textsl{EarlyLongSym}, which bracket the measurement of $\tau_e \approx 0.054$.

\subsubsection{Nyx}\label{ssec:nyx}
For our simulations we use the open source code\footnote{\url{http://github.com/AMReX-Astro/Nyx}} \texttt{Nyx}, a highly parallel, adaptive mesh, finite-volume N-body compressible hydrodynamics solver for cosmological simulations~\citep{almgren13,sexton21}. \texttt{Nyx} scales well on CPU- and GPU-based machines, and has been used extensively in studies of the IGM and Ly$\alpha$ forest~\citep[see e.g.][]{lukic15,onorbe19,chabanier23,wolfson23a,jacobus23}. We use a domain of 20 \mpchs with spatial resolution 10 \kpchs, which is higher than typically used in IGM studies, as we have found this to be necessary for convergence at $z>5$ given our numerical setup~\citep{doughty23}. 

We generate the transfer functions using {\sc CAMB}~\citep{lewis00}, and the initial conditions using the code CICASS~\citep{oleary12}, including a $\langle \Delta v \rangle =30$ km/s streaming velocity between baryons and dark matter at recombination. We have found that inclusion of a non-zero streaming velocity is found to create differences on the few per cent level in Ly$\alpha$ power, in particular leading to slightly reduced small scale power with respect to a $\langle \Delta v \rangle =0$ km/s simulation, also generated via CICASS. This difference increases with resolution, but down to at least $\sim\Delta x=2$ \kpchs is subdominant to other effects. Initial conditions are generated for $z=200$, and used to inform the overdensity field in AMBER as described in Section~\ref{sssec:amber}. We assume a $\Lambda$CDM cosmology consistent with results from Planck~\citep{planck20}, $\left(\Omega_M, \Omega_\Lambda, \Omega_b, h, X_H\right)=$ (0.315, 0.685, 0.049, 0.675, 0.76). 

\texttt{Nyx} excludes galaxy formation physics and we omit any type of feedback implementation, which should have a minimal effect on the high-$z$ Ly$\alpha$ forest~\citep{viel13a}, so the main impacts on the gas come from structure formation and the associated shock heating, and the reionization event itself. To model inhomogeneous reionization in \texttt{Nyx}, we use the coarse-grid reionization fields generated from AMBER, which are mapped onto the finer hydrodynamics grid. Once a hydrodynamics grid cell reaches its reionization redshift, a user-defined amount of heat is injected dependent on the current gas temperature and the self-consistently calculated hydrogen neutral fraction. 
\begin{equation}
    T^\mathrm{post-reion} = \left( x_\mathrm{HI}^\mathrm{pre-reion}\right) \left(\Delta T_\mathrm{re} - T^\mathrm{pre-reion} \right) + T^\mathrm{pre-reion}
\end{equation}
where $T^\mathrm{pre-reion}$ and $x_\mathrm{HI}^\mathrm{pre-reion}$ are the pre-reionization temperature and neutral fraction for a given cell. It can be heated by a possible maximum of $\Delta T_\mathrm{re}$, either 10000 (cold) or 20000 K (hot) in this work. We output simulation snapshots in bins of $\Delta z=0.1$ from $5.0 \leq z \leq 6.0$, and use these for all the analyses presented here.

We generate physical and Ly$\alpha$ optical depth skewers cast along all axes of the simulation, and in the positive or negative direction. The optical depths are calculated assuming that they are optically thin and in thermodynamic equilibrium, and do not use the self-consistent neutral fraction from \texttt{Nyx}. In theory, this could mean we would retrieve a non-zero flux from a cell that has not yet been reionized, which is not realistic. To address this, we set the H I photoionization rate to an extremely low value in cells that have not reionized by the snapshot redshift, thereby preventing any undue transmission.

\begin{figure*}
\includegraphics[width=\textwidth]{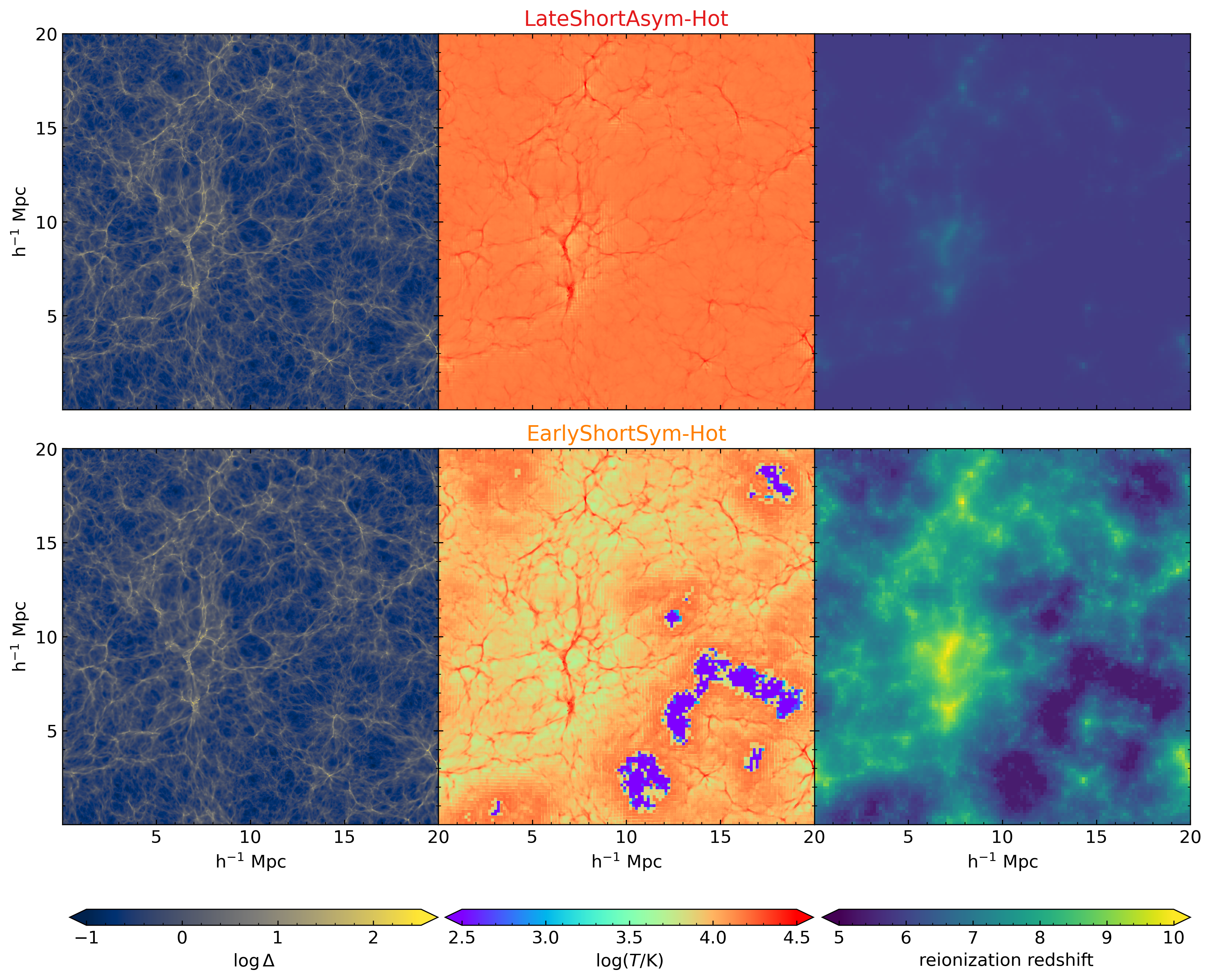}
\caption{50 km/s slices of a 20 \mpch, $2048^3$ simulation with a $128^3$ reionization field. The left column shows the gas overdensity field, while the center and right columns show the temperature and reionization redshift fields, respectively. Since the simulations have initial conditions with the same seed, the overdensity structures are the same; the differences come from the reionization histories. The top row shows the $\Delta T_\mathrm{re}$ versions of \textsl{LongShortAsym} and \textsl{EarlyShortSym} (red and orange curves in Figure~\ref{fig:xHI_plot}), with \textsl{LongShortAsym} being rapid with a late midpoint, and \textsl{EarlyShortSym} being prolonged with an early one. The rapid \textsl{LongShortAsym} history gives a homogeneous-looking $z_\mathrm{re}$ field, and a correspondingly featureless temperature field.} 
\label{fig:fiducial_slices}
\end{figure*}

\section{Results}\label{sec:results}
In this work, we are attempting to examine the trends in several Ly$\alpha$ opacity-based metrics of reionization, the 1D flux power spectrum, autocorrelation function, and effective optical depth distribution. To facilitate an understanding of those trends, we first characterize the impact of the differing reionization histories on the density and temperature fields, phase-space distribution of gas, and the temperature at mean density.

\subsection{Measures of gas in the simulation}\label{sssec:gas_in_sim}
\subsubsection{Slices}
In Figure~\ref{fig:fiducial_slices} we show 50 km/s slices in overdensity and temperature at $z=5.5$, and the reionization field for two reionization models: on the top row is the late, rapid, asymmetric, and hot (\textsl{LateShortAsym}) model, while on the bottom is an early, rapid, symmetric history, hot model (\textsl{EarlyShortSym}). For both models, overdense structures are visible in the left column as lighter colors, revealing the filamentary structure of the cosmic web. The temperature panels (center column) show hotter temperatures ($>10^{4.5}$ K) co-located with the filaments, which trace collapsed structures as they have been shock heated to their virial temperatures. Due to the identical seeds used in the initial conditions, the same overdensity structure is naturally present in both models.

Beyond these commonalities, the differing natures of the two histories become apparent. The abrupt \textsl{LateShortAsym} model shows a very uniform reionization field (right column), with only the \emph{most} overdense regions ionizing at barely over $z=6$, while \textsl{EarlyShortSym} has a much wider dynamic range in $z_\mathrm{re}$ values. Accordingly, \textsl{LateShortAsym} has a relatively featureless temperature field in the underdense IGM, nearly all cells hovering around $\log \; (T\mathrm{/K}) \sim4.2$. There are minor excursions to lower temperatures ($\sim10^{4.0}$ K) around the largest filaments, for example near (8, 8) \mpchs, but they are quite minimal and difficult to pick out. For \textsl{EarlyShortSym} on the other hand, the regions surrounding the filaments are those that reionized first, indicated by yellow coloring in the reionization field in the rightmost column. Since those regions became reionized first, they have had the most time to cool, and have achieved much lower temperatures than the regions which were more recently ionized. This slice also captures a few completely un-ionized regions in the \textsl{EarlyShortSym} simulation, which doesn't fully complete (in a volume-weighted sense) until $z=5.4$; these are the glaringly cold purple/blue cells visible in the lower center panel of Figure~\ref{fig:fiducial_slices}. Since they have not been photoheated by reionization and are physically distant from structure formation-induced shock heating, they have been permitted to cool adiabatically since recombination, reaching temperatures of $<100$ K.

\subsubsection{Temperature density relation}
\begin{figure*}
\includegraphics[width=\textwidth]{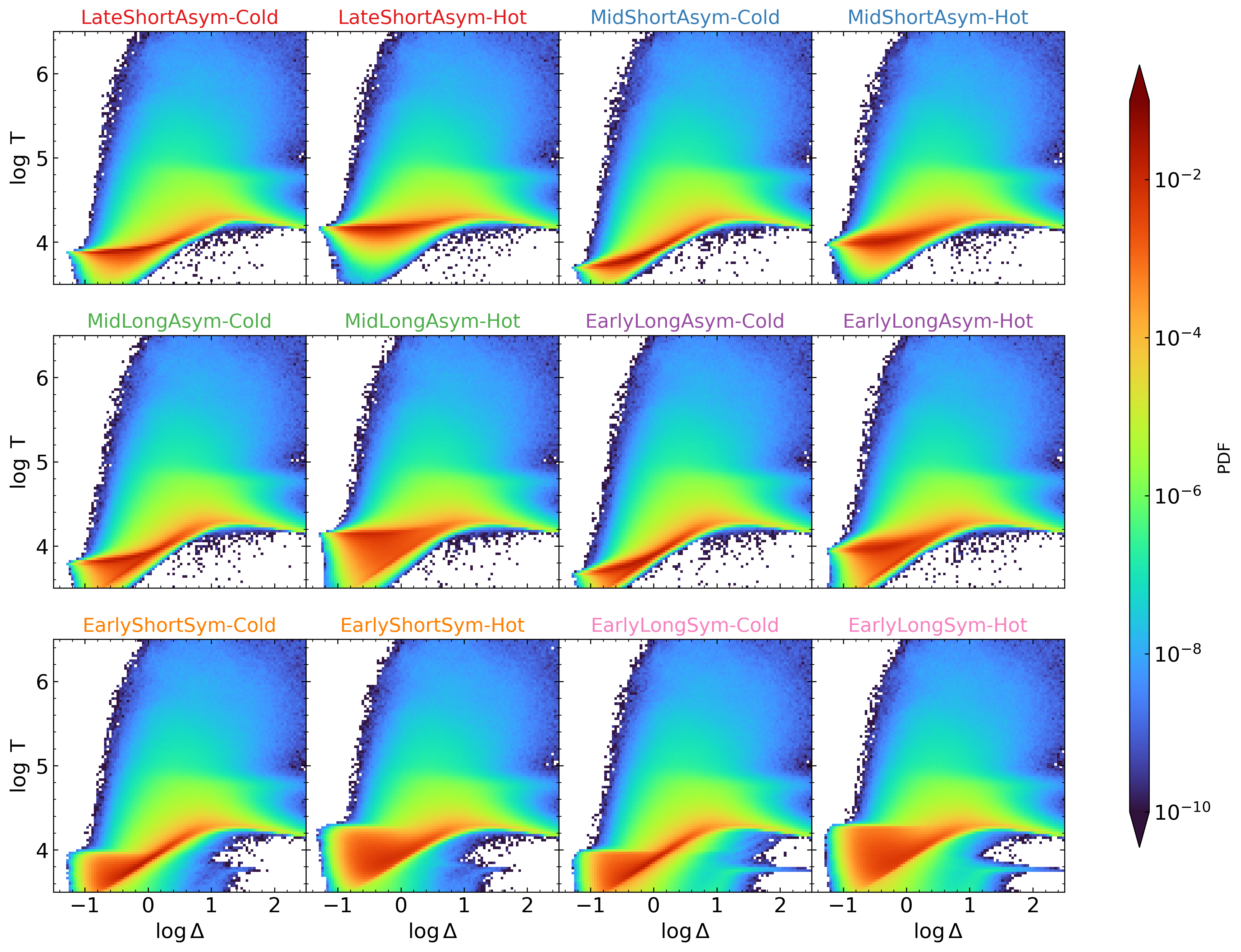}
\vspace{-0.2cm}
\caption{
Temperature density relations of all 12 models, grouped based on visual similarity into rows, from top to bottom: Short asymmetric (\textsl{ShortAsym}), long asymmetric (\textsl{LongAsym}), and symmetric (\textsl{Sym}). From this visualization, it is apparent that an earlier reionization midpoint allows the gas to further relax, and move away from a flat isothermal relation (e.g. \textsl{LateShortAsym} versus \textsl{MidLongAsym}). Asymmetry causes a large injection of heat at later times, more resembling an instantaneous reionization, while a symmetric early history permits more gas to have relaxed into the expected temperature-density relation.}
\label{fig:rhot}
\end{figure*}
It is expected that gas in the IGM will settle into a relatively tight relation between temperature and density after reionization has completed~\citep{hui97}
\begin{equation}\label{eq:temperature_density_relation}
    T(\Delta) = T_0 \Delta^{\gamma-1}
\end{equation}
where $\Delta$ is the overdensity, $\rho/\langle \rho \rangle$, and $\gamma$ is the adiabatic index, found to be $\sim1.6$ in a relaxed medium. For an inhomogeneous reionization, however, the amount of time since reionization occurred will vary between different regions in the simulation, meaning that only a fraction of the gas will obey the relation, and instead have a wider overall spread in properties~\citep{onorbe19}. To examine the impact of the reionization parameters on the distribution, we plot the temperature-density distributions for all models at $z=5.5$ in Figure~\ref{fig:rhot}. 

For the short asymmetric histories (\textsl{ShortAsym}, top row), the bulk of the gas occupies a relatively narrow range of temperatures for a given overdensity, although this is not yet the relaxed temperature-density relation of Equation~\ref{eq:temperature_density_relation}. All else being equal, models with higher heat injection have a flatter distribution, resembling that obtained when assuming an instantaneous reionization, whereas the colder models have moved closer to their eventual log-log distribution. Long, asymmetric histories (\textsl{LongAsym}; second row) tend to have gas split between the final $\Delta-T$ relation with steeper $\gamma$ and a flatter, unsettled distribution more resembling the \textsl{ShortAsym} models. These are visible as two dark red lines at $\log \Delta < 1$.

Symmetric models (\textsl{Sym}; bottom row) create two distinguishing features in their distributions. First, there is gas remaining below the temperature density relation, whereas for all the other models this region of the phase space is essentially empty. This gas, still mostly at IGM overdensities, is removed by the reionization process, but these reionization histories end the latest, both after $z=5.5$.\footnote{There is a ``hook'' of gas that first increases logarithmically from $(\log \Delta, \log (T/\mathrm{K})=(1.0, 3.5)$ before becoming flat out to $\log \Delta >1.4$. This is likely shock heated gas that in the other simulations has already been removed from this phase by additional heating reionization process.} Further, these models resemble the \textsl{LongAsym} ones in that they have a wider range of temperatures below $\log \Delta = 1$ than the short histories. However, more of their gas has settled into the final log-log temperature-density relation, and they are lacking material in the upper dark red line present in the \textsl{LongAsym} histories. As seen in Figure~\ref{fig:xHI_plot}, it seems that asymmetric histories contribute to the majority of the gas being reionized in a single, impulsive event, causing a higher temperature and flatter $\rho-T$ relation prior to the IGM's thermodynamic relaxation.

\subsubsection{Temperature at mean density}
\begin{figure}
\hspace*{-0.3cm}\includegraphics[width=0.5\textwidth]{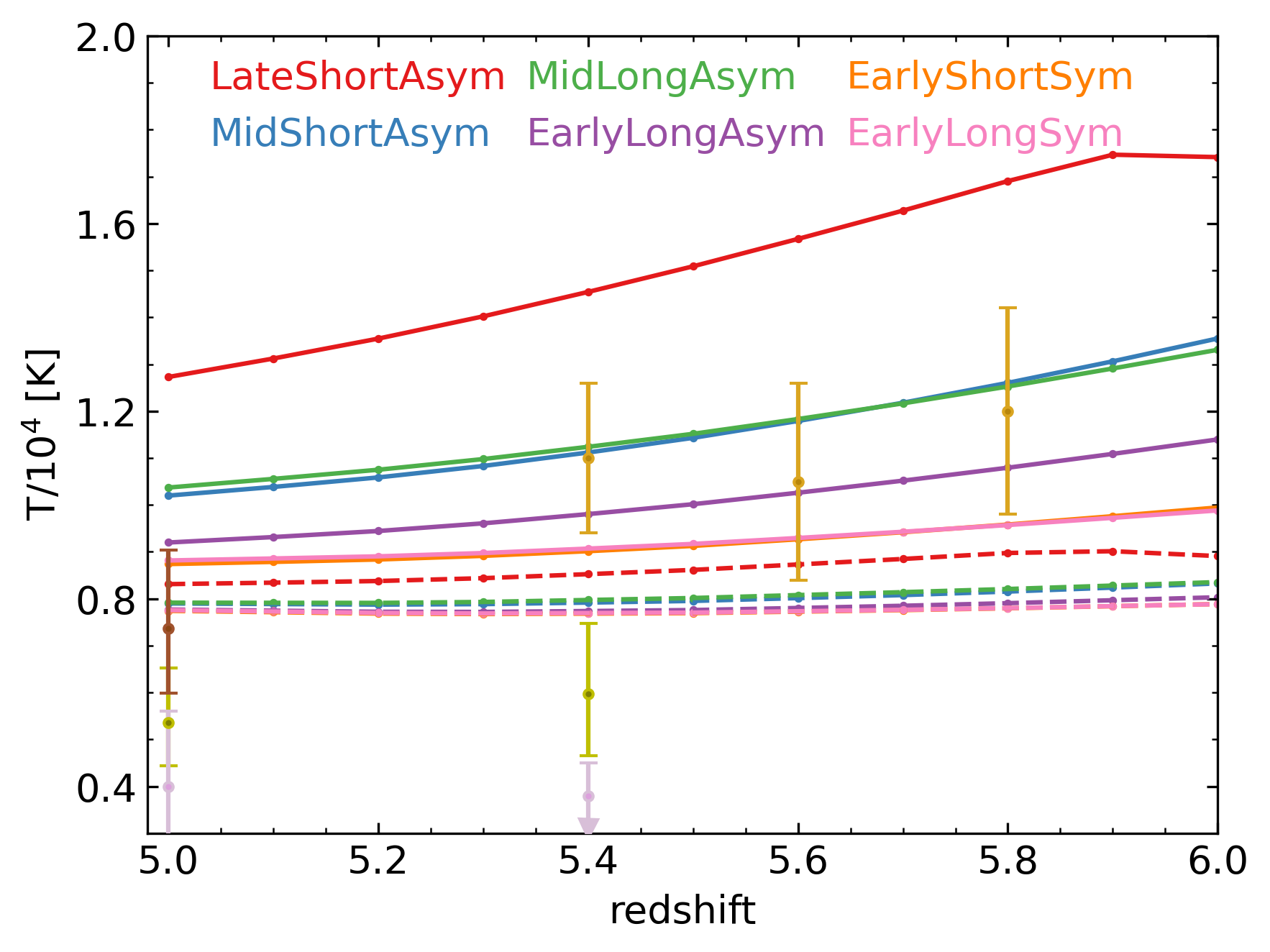}
\caption{The temperature at mean density for our simulation runs, overplotted with values derived from $z\geq5$ observations~\citep{garzilli17,boera19,walther19,gaikwad20}. The hot models are indicated with solid lines, while dashed lines are for the cold ones. The model trajectories span a wide range of temperatures, but are not well-constrained by the observations, which are themselves in contention with one another.}
\label{fig:T0_plot}
\end{figure}

An important part of Equation~\ref{eq:temperature_density_relation} is the temperature at mean density, $T_0$, as it is reflective a combination of heating and cooling events impacting the majority of gas in the Universe. We calculate the temperature at mean density from $5 < z < 6$ for each of our simulation runs, and show them in Figure~\ref{fig:T0_plot}. These are calculated from the full snapshot data outputted at each $\Delta z=0.1$ interval redshift, within the overdensity range $0.95 < \Delta < 1.05$. As a point of comparison, we overplot observationally-derived temperatures from several studies~\citep{garzilli17,boera19,walther19,gaikwad20}. Since a nonzero Ly$\alpha$ signal can't be obtained from the un-ionized IGM, values derived from observations are probing reionized regions, so we also restrict contributions to cells that have been ionized by masking out those with $z_\mathrm{re}<z_i$. In practice, this is only relevant for models with $z_\mathrm{end}<6.0$.

First, we find that all of the cold models (dashed lines) have very similar temperatures, $\sim$8000 K, and do not show much cooling over $5 < z < 6$, instead being relatively constant. The hot models all show greater temperature evolution, and the inter-model variation covers a much wider range of temperature. We find that our extreme \textsl{LateShortAsym} model creates a $T_0$ that is too high by a factor of $\sim1.5$, lying above even the relatively high derived temperatures in~\citet{gaikwad20}. The other hot models \textsl{MidShortAsym}, \textsl{MidLongAsym}, and \textsl{EarlyLongAsym} all cut through the~\citet{gaikwad20} datapoints, with \textsl{MidShortAsym} and \textsl{MidLongAsym} overshooting the $T_0<10000$ K measurements from other studies. \textsl{EarlyShortSym} and \textsl{EarlyLongSym} lie the closest to the cold model trajectories, approximately passing between the low $T_0$'s from~\citet{garzilli17} and~\citet{walther19} and the high values from~\citet{gaikwad20}, and matching the $z\sim5$ data point from~\citet{boera19}.

It is interesting to note the cases where different reionization histories have induced very similar $T_0$ trajectories. The \textsl{MidShortAsym} and \textsl{MidLongAsym} histories have nearly identical $T_0$ evolution, with the \textsl{MidLongAsym} history being more ionized at $z>6.6$ and completing about $\Delta z=0.5$ later, while having a similar midpoint. The two \textsl{Sym} histories are even more similar. Evidently, despite \textsl{EarlyLongSym}'s extremely late end, the continued heat injection at late times is from such a small number of cells ($x_\mathrm{HI,v}(z\sim5.4)=0.05$), that it doesn't matter for the average temperature. With the exception of the hot \textsl{EarlyLongAsym} model, the similar histories are ones with similar mass-weighted midpoints.

\subsubsection{Pressure smoothing scale}
The instantaneous temperature of the IGM is important for understanding its structure, but in actuality it takes time for the gas to thermodynamically respond to a heating event like reionization. To quantify the physical response at a given redshift, we examine the pressure smoothing scale, also called the filtering scale~\citep{gnedin98}, describing the physical scale below which there is a cutoff in smaller structures. This is due to pressure smoothing induced by the instantaneous temperature and also the thermal history. 

To do this, we first calculate the 3D matter density power spectra in all our redshift bins, restricting it to the gas at overdensities $0.2 < \Delta < 0.4$ to isolate the underdense gas contributing to the Ly$\alpha$ forest. We then compare the 3D power $P_\mathrm{rad}(z)$ to that in an otherwise identical simulation that has been allowed to adiabatically cool with no reionization event, and thus excludes any pressure smoothing effect, $P_\mathrm{ad}(z)$. The expected relationship is characterized as
\begin{equation}\label{eq:pressure_smoothing}
P_\mathrm{rad}(k) = N \cdot P_\mathrm{ad}(k) \cdot \exp (-k^2/k_\mathrm{ps}^2)^2
\end{equation}
as defined in ~\citet{puchwein23}~\citep[and matching our previous treatment in][]{doughty23}. We then fit for a normalization parameter $N$ and the pressure smoothing wavenumber $k_\mathrm{ps}$. 

\begin{figure}
\hspace{-0.5cm}
\includegraphics[width=0.5\textwidth]{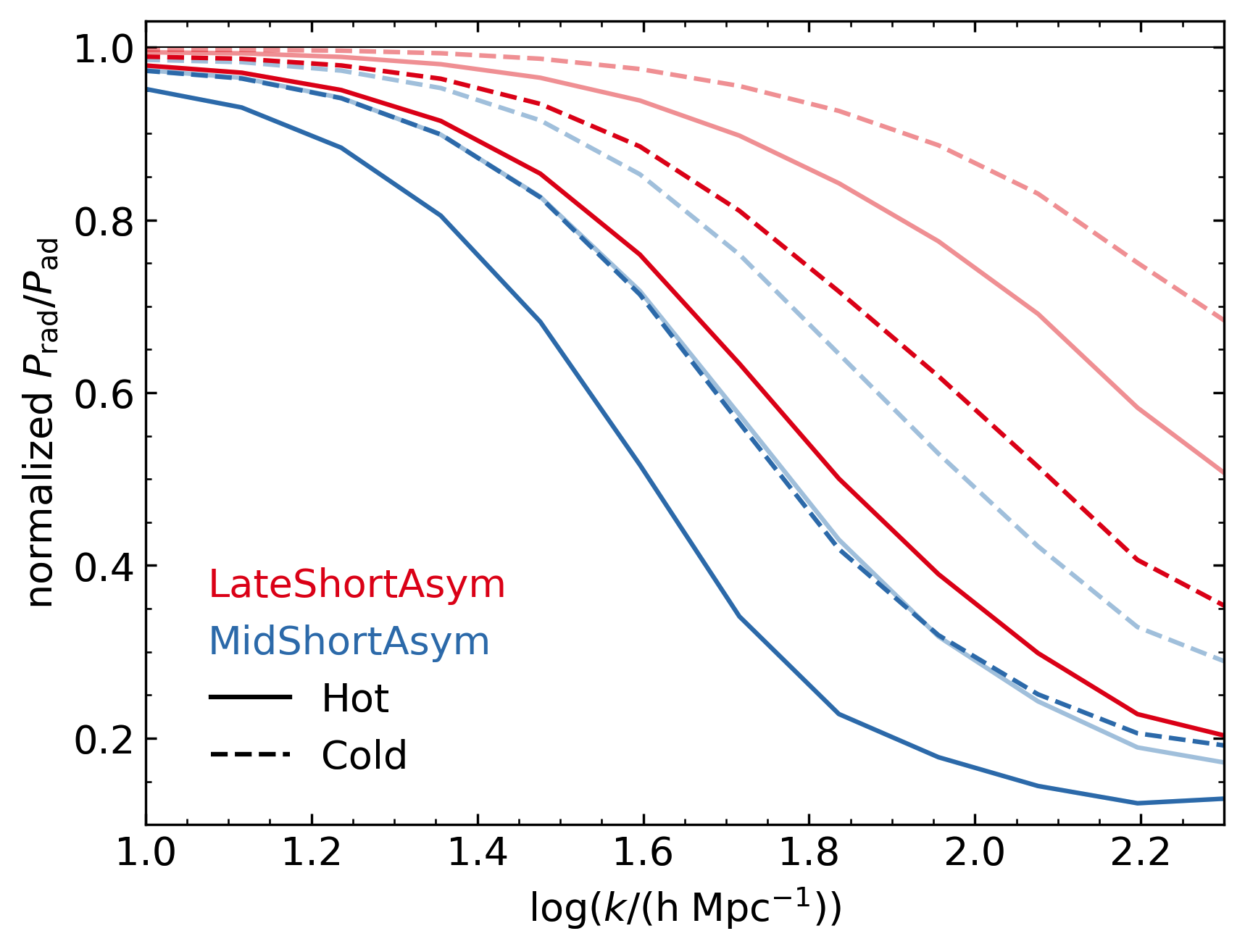}
\caption{The normalized ratios of Equation~\ref{eq:pressure_smoothing} showing the loss of small scale power due to heat-induced pressure smoothing, using the \textsl{ShortAsym} models as an example. The lines with lower (higher) saturation are taken at $z=5.5$ ($z=5.0$), showing how the smoothing effect is lower at higher $z$. \textsl{LateShortAsym} and \textsl{MidShortAsym} are identical except for their reionization midpoint, and the earlier midpoint in \textsl{MidShortAsym} leads to increased smoothing when considered at the same redshift. Hot models (solid lines) also are more smoothed than cold ones (dashed).}
\label{fig:pressure_smoothing}
\end{figure}

We plot the normalized ratios $P_\mathrm{rad}/P_\mathrm{ad}$ for both of the \textsl{ShortAsym} models as an example in Figure~\ref{fig:pressure_smoothing}. We include both the hot and cold versions (solid and dashed lines) and at $z=5.5$ and $z=5.0$ (less and more saturated in color, respectively). Models that are more removed from the $\mathrm{ratio}=1$ line have more pressure smoothing, i.e. they have less power on small scales and so fall off farther to the left.

\begin{figure}
\hspace{-0.3cm}
\includegraphics[width=0.48\textwidth]{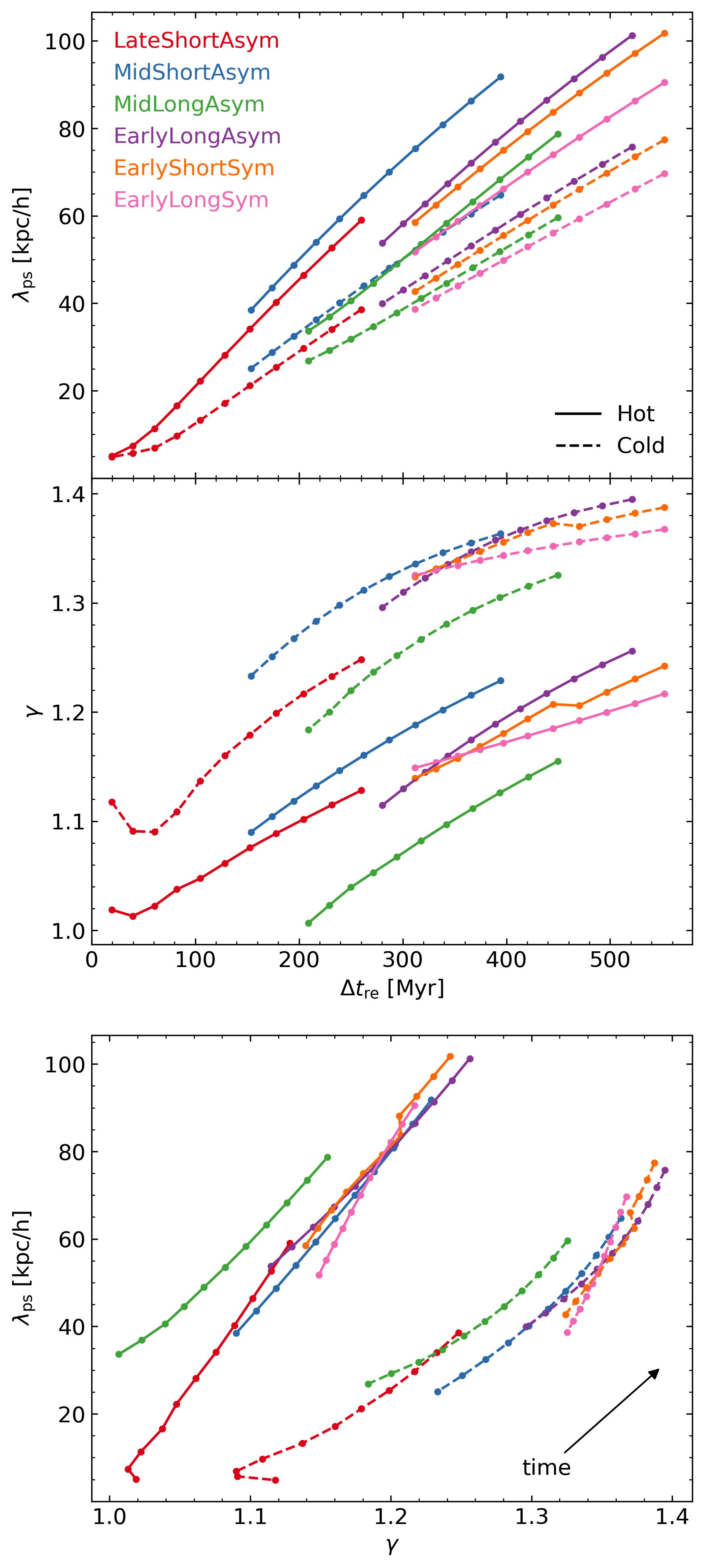}
\caption{\emph{Top and middle}: The evolution of the pressure smoothing scale (top) and the measured adiabatic index (middle) with time since the mass-weighted reionization midpoint. The points indicate times corresponding to each of our redshift snapshots. Since both are related to thermodynamic relaxation, they both increase with time. \emph{Bottom}: The correlation between the measured pressure smoothing scale and the adiabatic index, measured in the same simulation snapshots. Both values increase with time, and so are strongly correlated with one another for a given model. Many of the model trajectories overlie one another, but there is an obvious separation between the hot and cold models (solid and dashed lines, respectively).}
\label{fig:lps_gamma}
\end{figure}

\begin{figure*}
\hspace{-0.5cm}
\includegraphics[width=\textwidth]{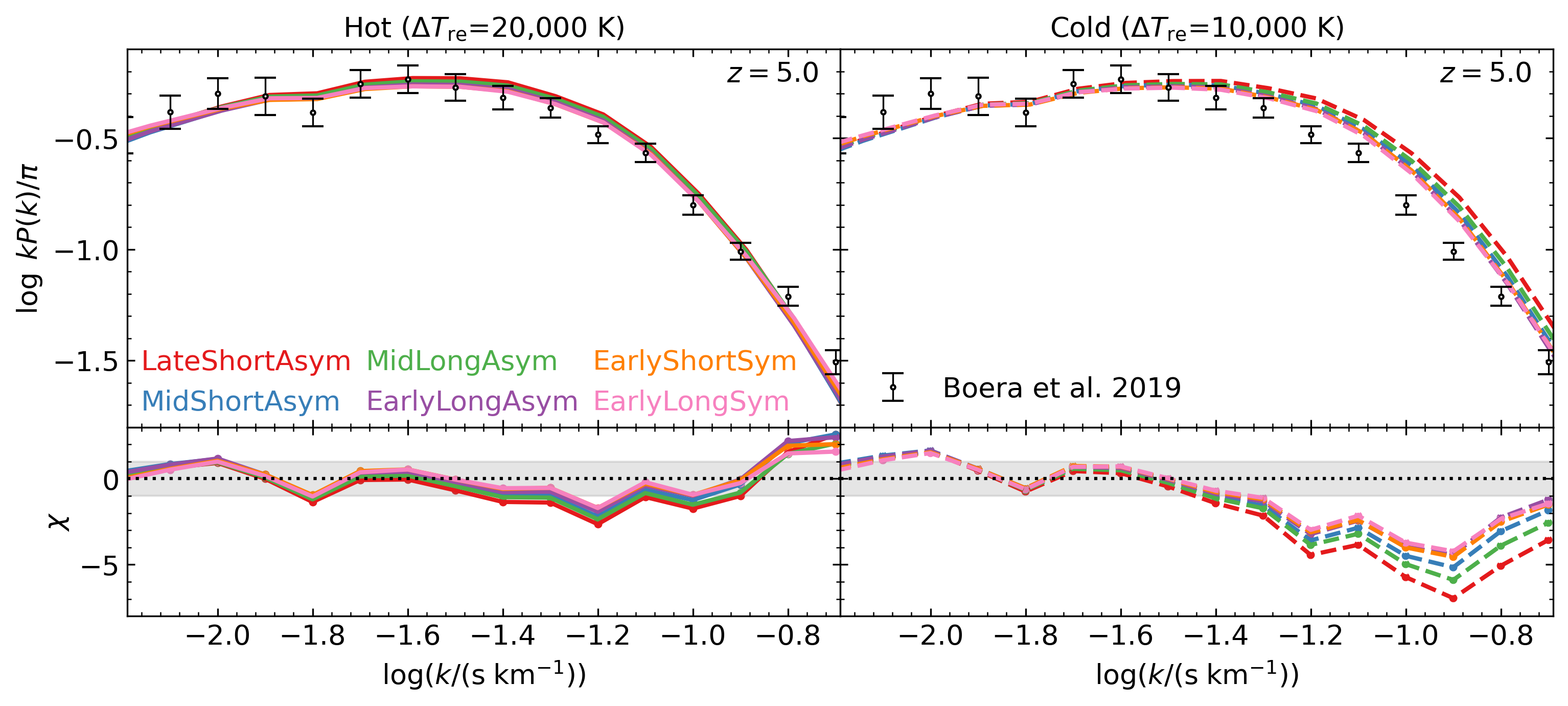}
\caption{Power spectra of 1D Ly$\alpha$ flux, with data from~\citet{boera19} at $z=5$. All hot models (left panel) have extremely similar values and generally lie quite close to the observed values, while the cold models (right panel) differ from each other, and demonstrate excessive small scale power. For the cold models, the order of increasing high $k$ power matches the order of increasing temperature at mean density. This is likely because the temperature is an indicator of time since the reionization midpoint, and a more recent midpoint results in a smaller pressure smoothing scale.}
\label{fig:Lya_power}
\end{figure*}

There are trends here which can serve as a reality check on the method, by confirming our intuition about how the pressure smoothing scale should evolve with time and temperature. First, the hot versions of each reionization history peel away from the $\mathrm{ratio}=1$ line at larger physical scales than their cold counterparts, which makes sense given the increased speed of sound in hotter gas, and the consequently increased ability for the gas to expand post-reionization. The history with the $z\sim6$ midpoint \textsl{LateShortAsym} shows less pressure smoothing than  the earlier one, which is sensible given that it has had less time to physically respond to the heat injection. Lastly, of course the pressure smoothing is larger for any given model at $z=5$ than at $z=5.5$, again since more time has passed since the thermal injection event.

Given that both the pressure smoothing scale and the slope of the $\rho-T$ relation $\gamma$ are related to the time elapsed post-reionization, we examine the relationship between these two quantities. Fitting the $\rho-T$ distributions in Figure~\ref{fig:rhot} we extract the adiabatic index $\gamma$ and plot both it and our measured pressure smoothing scale $\lambda_\mathrm{ps}$ (equal to $2\pi/k_\mathrm{ps}$) against the time elapsed since the mass-weighted reionization midpoint, $\Delta t_\mathrm{re}$. The results are plotted in Figure~\ref{fig:lps_gamma}.

The fitted pressure smoothing scale curves (top panel) start where $\Delta t_\mathrm{re}$ is the time difference between the mass-weighted midpoint and the time at $z=6$, and increase approximately linearly with time. Extrapolating the curves to $\Delta t_\mathrm{re}=0$, they radiate from some initial smoothing scale $\lesssim5$ \kpch. Comparing between hot and cold versions of the same reionization histories, the hot models have steeper slopes and have larger values for a given $\Delta t_\mathrm{re}$. The general linear trend is supportive of the simple expansion model shown in~\citet{puchwein23}.

The time evolution of the adiabatic indices is a bit messier, as while they increase with time the rate of change slows and there is a turnover at higher $\gamma$. This is particularly evident in the cold models, which also start at higher values than their hot counterparts, a logical result given that it does not take as long for a model with low heat injection to thermodynamically relax into the expected power-law relation. Unlike the pressure smoothing scale, these curves do not all emanate from the same point; however, we should expect that the ``initial'' $\gamma$ would be close to one, particularly for the \textsl{Asym} models with their quasi-instantaneous reionizations and resultant nearly isothermal temperature distributions.

We also directly compare $\lambda_\mathrm{ps}$ and $\gamma$ measured at the same redshifts in the bottom panel of Figure~\ref{fig:lps_gamma}, with the hot models indicated with circles and the cold models with squares. Of course, the values both increase with decreasing redshift (lower left to upper right within the panel), and the hot and cold models occupy different regions of the plot, with cold models having higher $\gamma$'s for a given $\lambda_\mathrm{ps}$. Many of the $\lambda_\mathrm{ps}-\gamma$ curves even lie on top of one another, particularly for the hot models, even for significantly different reionization histories. Based on these results, in the Ly$\alpha$ metrics we should expect to see evidence of more small scale structure in the cold models, even if they are more thermodynamically relaxed.

\subsection{Metrics of the Ly$\alpha$ forest}\label{ssec:gas_in_laf}
\subsubsection{1D Ly$\alpha$ forest power spectrum}
The 1D Ly$\alpha$ flux power spectrum characterizes the fluctuations in the flux field as a function of the wavenumber $k$
\begin{equation}
P_\mathrm{Ly\alpha}(k) = \int d^3 \bm{x} \, \xi (x) \, e^{-i \, \bm{k} \, \bm{x}}
\end{equation}
where $\xi (\bm{x})$ is the two-point correlation function, or autocorrelation function
\begin{equation}
    \xi (x) = \langle \delta_F(\bm{x}_1) \, \delta_F(\bm{x}_2) \rangle 
\end{equation}
with $x=\lvert \, \bm{x_1}-\bm{x_2} \, \rvert$ and $\delta_F=F/ \langle F \rangle - 1$. Greater power indicates greater fluctuations in the flux contrast on that spatial scale, and thus a larger amount of structure with size $\lambda= 2 \pi/k$. We calculate the power for our models at $z=5$, and plot them in Figure~\ref{fig:Lya_power} alongside the measurements from~\citet{boera19} who were able to probe $-2.2 \leq \log$\skm$\leq -0.7$, reaching an unprecedented high $k$ mode achievable through use of high S/N spectra. All models have had their optical depths adjusted to produce a mean flux of $\langle F \rangle = 0.18$, matching the observations.

We find that all of the hot models (left panel) have extremely similar power spectra, with the largest inter-model differences not exceeding the observational uncertainties. Additionally, they are all within \emph{approximately} one $\sigma$ of the observed values for the entire $\log k$ range, except for the bins at $\log$\skm$=-1.2$ and $\log$\skm$\geq-0.8$. The cold models (right panel) differ more from one another than the hot ones, especially for higher $k$, with the largest difference amounting to $\sim 3 \sigma$ at $\log$\skm$=-0.9$. As noted earlier, all of the cold models have relatively low $T_0$, but the ordering of the trend in small scale power is interesting, because the models with higher power are those with slightly higher temperature at mean density. Thus, for cooler reionization models it doesn't seem that the temperature difference is the driver of the differences. It may instead be that a slightly higher temperature here is more indicative of a later or ongoing reionization, meaning the small structure of the IGM has had less time to react to reionization-induced heating, resulting in a smaller pressure smoothing scale. 

Also interesting is the shape of the cold models, which compared to the hot models show a steeper average slope for $\log$\skm$<-1.6$ and a flatter slope for $\log$\skm$>-1.4$. These models clearly do not create the correct shape in the power spectrum to match~\citet{boera19}. However, at the same time, the hot models don't produce sufficient $\log$\skm$>-0.9$ power to match the observations, and both the hot and cold models may be too steep at $\log$\skm$<-1.8$. The latter could potentially be alleviated by introducing UVB fluctuations, which would preferentially impact large scales~\citep{onorbe19}. For the mismatch at small scales, perhaps an intermediate injection temperature, $\approx$15000 K, is necessary.

\begin{figure*}
\includegraphics[width=\textwidth]{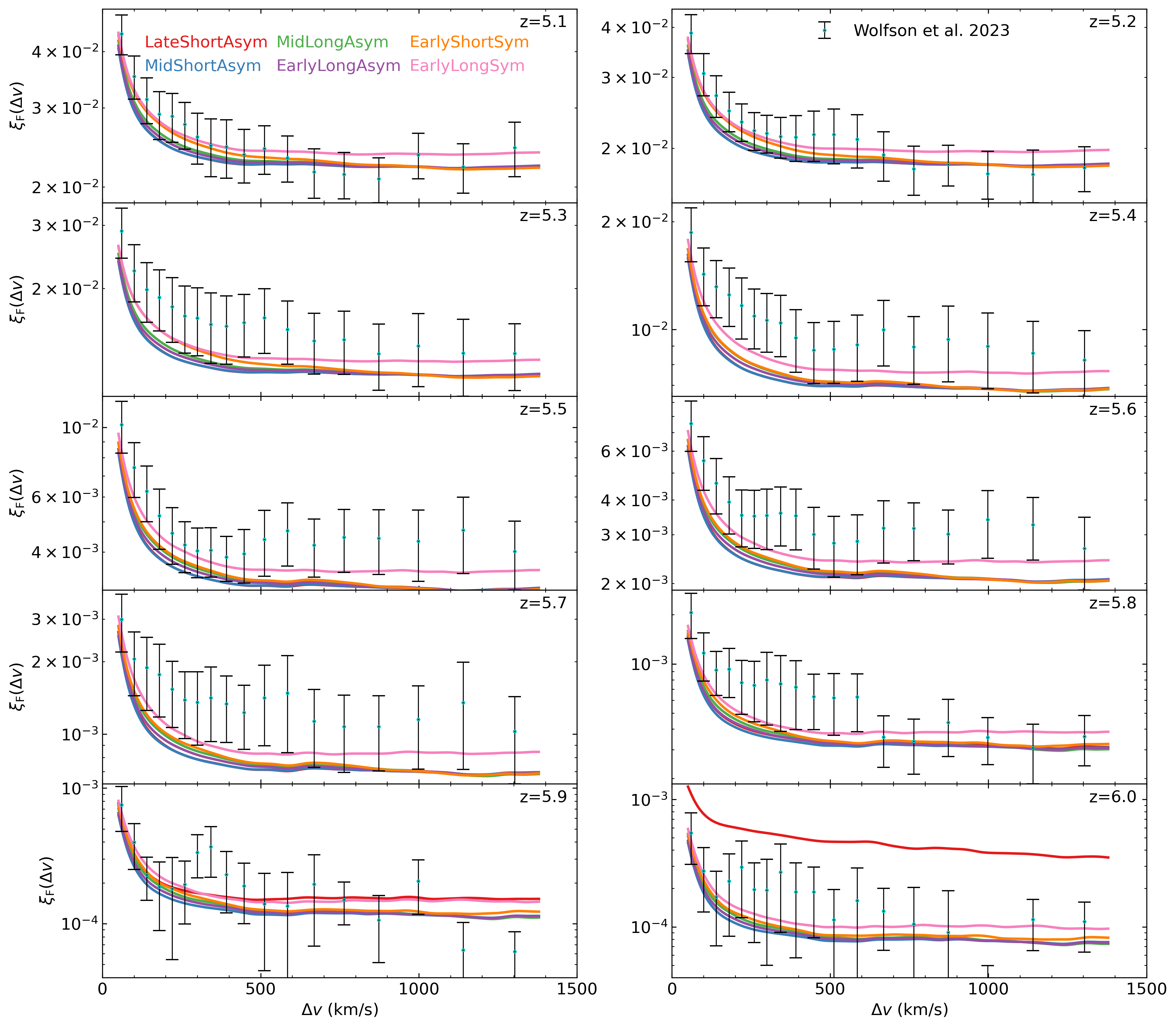}
\caption{The flux autocorrelation functions from $z=5.1$ to $z=6.0$ overplotted with the results from~\citet{wolfson23b}. The mean fluxes of each skewer set have been adjusted to match the observed quantities in each $\Delta z=0.1$ redshift bin. The simulations generally lie low compared to the observed points especially at smaller lags, $<700$ km/s. The exceptions are models that still have significant neutral material, such as \textsl{LateShortAsym} at $z=6.0$ and \textsl{EarlyLongSym} in all $z$ bins, although the latter is still systematically low.}
\label{fig:xi_hot}
\end{figure*}

\begin{figure*}
\includegraphics[width=\textwidth]{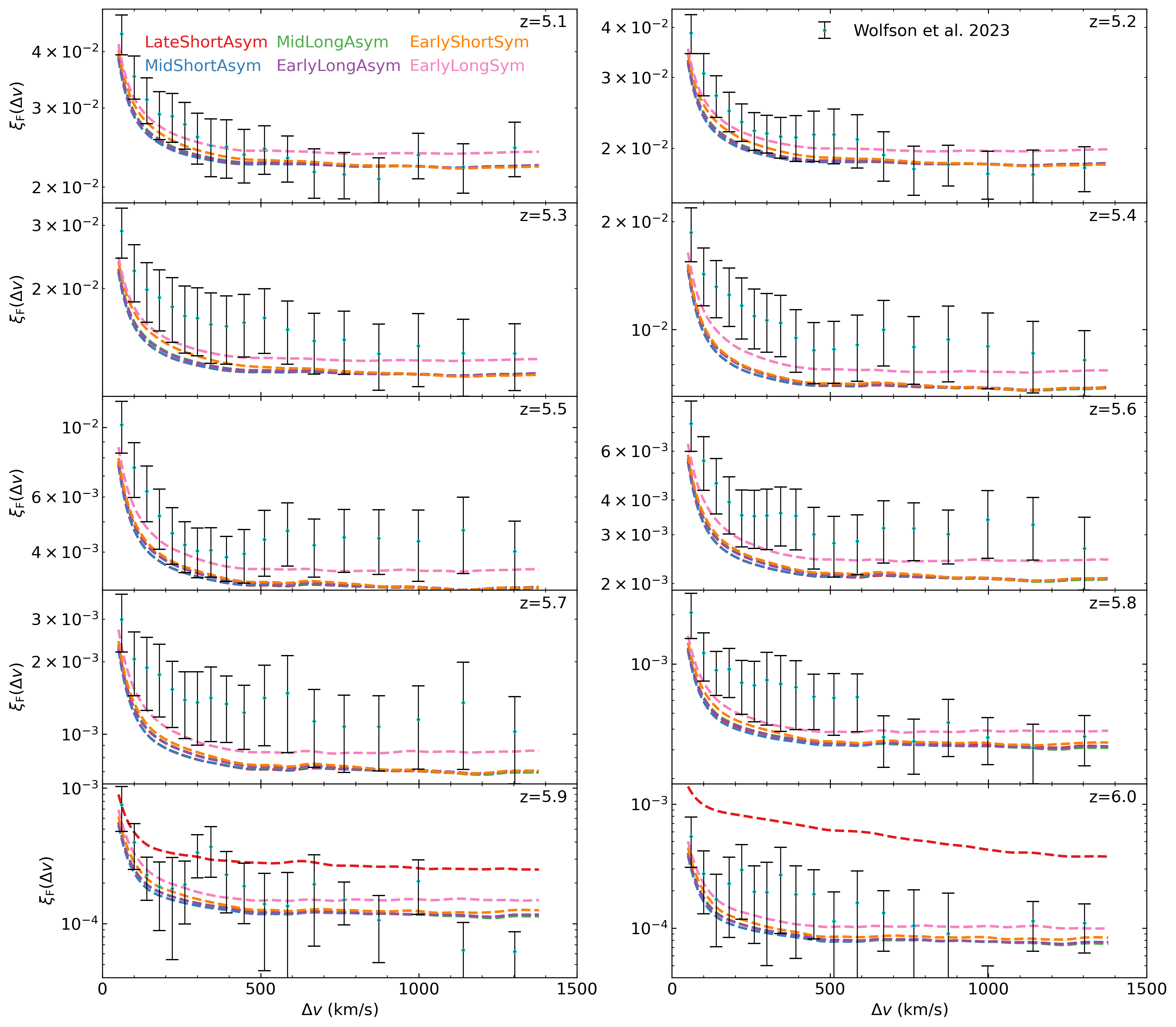}
\caption{The same as Figure~\ref{fig:xi_hot}, but for the cold models. The model differences here are smaller than for the hot models, likely as a result of reduced temperature fluctuations given the lower maximum heat injection at reionization. They are nearly identical to the hot models at lags $\Delta v>500$ km/s, but lower at smaller lags.}
\label{fig:xi_cold}
\end{figure*}

\subsubsection{Ly$\alpha$ autocorrelation function}
The flux autocorrelation function $\xi(x)$ goes into the calculation of the power spectrum, and so contains another representation of the same information. However, it is easier to measure than the power spectrum, and is less impacted by observational necessities such as masking. Therefore, it is useful to understand its sensitivity to the reionization history. Typically, the autocorrelation is given as the function of the velocity lag between pixels and calculated on the flux contrast. However, at $z>5$ when the mean flux is quite low, this can result in large vertical shifts in the function from a relatively small uncertainty in the mean flux, so we elect to calculate it on the raw flux field instead. For each raw model skewer, we first apply smoothing to account for the approximate instrumental resolution of X-Shooter, $R\sim8800$, and then calculate the function as
\begin{equation}\label{eq:autocorr_func}
    \xi_F \left( \Delta v \right) = \langle F(v) F(v+\Delta v) \rangle
\end{equation}
where $F(v)$ is the flux in the pixel at velocity $v$ and $\Delta v$ is the applied velocity lag. Since the simulation reinforces a periodic boundary condition at the domain edges, there is a close physical connection between pixels at opposite ends of each skewer, so we account for this in our calculation. We plot the results for the hot models from $5.1 \leq z \leq 6.0$ in Figure~\ref{fig:xi_hot} and the cold models in Figure~\ref{fig:xi_cold}. Plotted alongside are the results from the~\citet{wolfson23b} analysis of XQR-30 data, where we have matched our skewers to their measured mean fluxes for each $\Delta z$ bin.

With the exception of \textsl{LateShortAsym}, and the two \textsl{Sym} models, the values lie very close together in all redshift bins, showing the greatest inter-model deviations mostly at smaller lags. At $\Delta v>700$ km/s or so, nearly all the models converge to $\langle F \rangle ^2$, which is expected for $\xi_F$, whereas for the more standard $\xi_{\delta_F}$ it would converge to $\sim0.0$. \textsl{LateShortAsym} lies above the other models at $z=6.0$, but has settled by $z=5.9$. This rapid change can be explained by the continuing reionization of this model at these redshifts: it is still 60 per cent neutral at $z=6$ and does not finish reionizing until $z=5.9$, so large fluctuations in the field of neutral gas cause an elevated $\xi_F$ at $z=6$. The two \textsl{Sym} models both have higher $\xi_F$ than the others, and the more neutral of the two (\textsl{EarlyLongSym}) is consistently the highest for $z\leq5.9$. This persists down to $z=5.3$, at which point \textsl{EarlyShortSym} finishes reionizing and rises to match \textsl{EarlyLongSym} at $\Delta v<500$ km/s.

With regards to the data from~\citet{wolfson23a}, while the modeled autocorrelation values often fall within the 1$\sigma$ limits, they fall systematically low, particularly at $5.4 \leq z \leq 5.8$. The agreement is a bit better for $z\leq5.2$ and $z\geq5.9$, at the higher $z$ especially because the measurements are quite noisy. They also match the observations better at larger lags, where everything beings to converge to $\langle F \rangle ^2$.

The general trends in the cold models (Figure~\ref{fig:xi_cold}) are the same, but there are some differences with respect to the hot models. First, the cold $\xi_F$ values are slightly lower, except for \textsl{LateShortAsym} at $z\geq5.9$. Second, they are lower specifically at $\Delta v<500$ km/s, but converge to the same values as the hot models at larger velocity lags. It seems that the hot models are generating greater temperature fluctuations to elevate the overall variance and by extension the zero-lag autocorrelation function.

\subsubsection{Cumulative distribution of effective optical depths}
While the mean flux and the associated effective optical depth $\tau_\mathrm{eff}$ measured using quasar sightlines is a useful global statistic, there is more information contained within the distribution of mean fluxes along each individual line of sight, $\langle F \rangle_\mathrm{los}$. In particular, it is another way of characterizing the amount of scatter in the opacity, with a broader distribution reflecting greater variation in the opacities, and the position of the center of the distribution being reflective of the typical amount of saturation~\citep{fan23}.

\begin{figure*}
\hspace*{-0.1in}
\includegraphics[width=\textwidth]{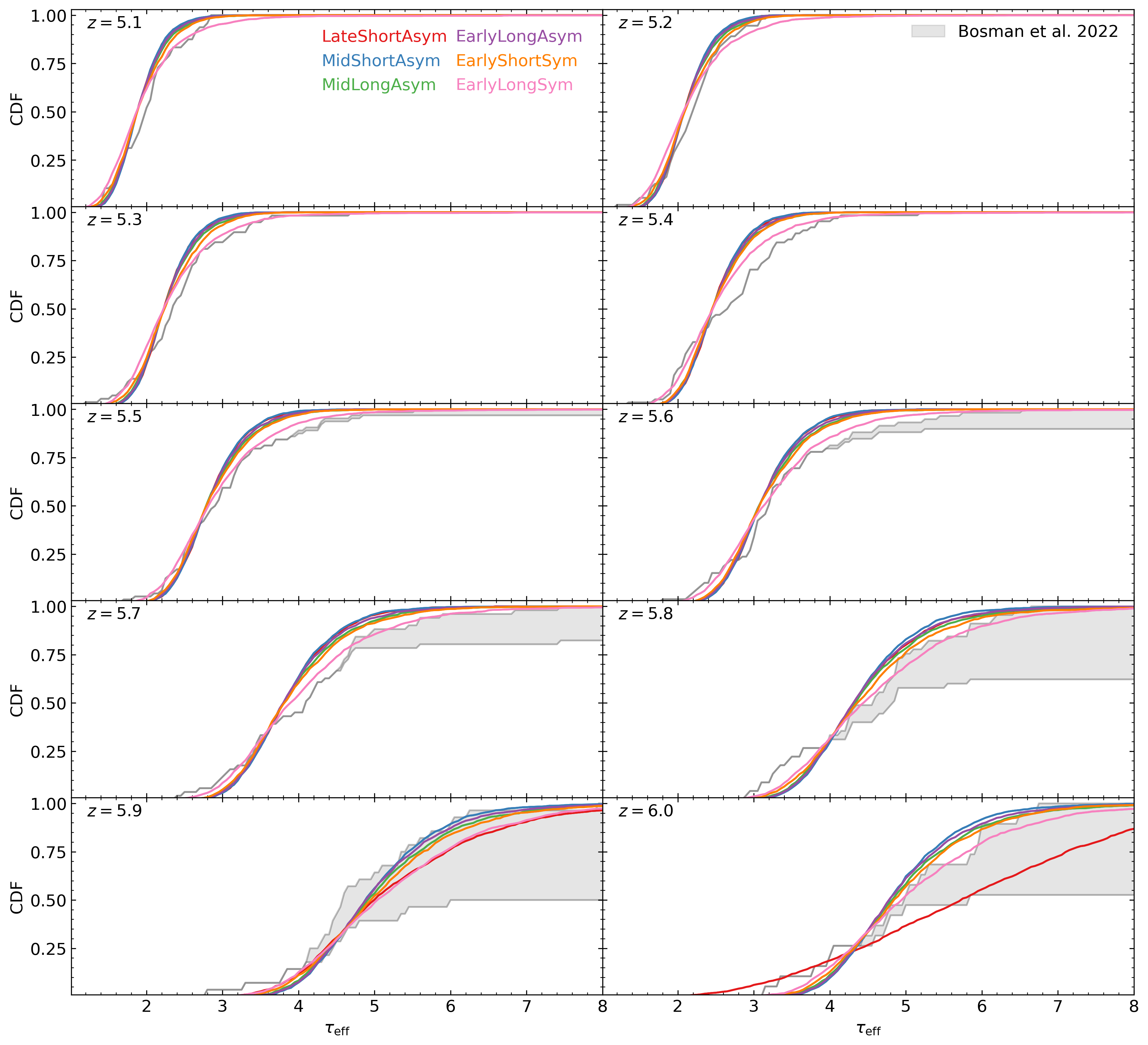}
\caption{Effective optical depth distributions for all models in $\Delta z=0.1$ intervals, plotted with observations using the data from~\citet{bosman22} reprocessed for segment lengths matching our domain size, obtained via private communication. The mean flux for each bin is set to the mean of all the observed lines of sight, because this does not match the reported mean flux in each redshift bin. The models are generally very similar to one another, except in the upper 40 per cent of the CDF, and show a narrower distribution than the data. The exception is the late-finishing \textsl{EarlyLongSym}, which generally matches the upper limits of the observed distribution.}
\label{fig:tau_eff_cdf_hot}
\end{figure*}

\begin{figure*}
\hspace*{-0.1in}
\includegraphics[width=\textwidth]{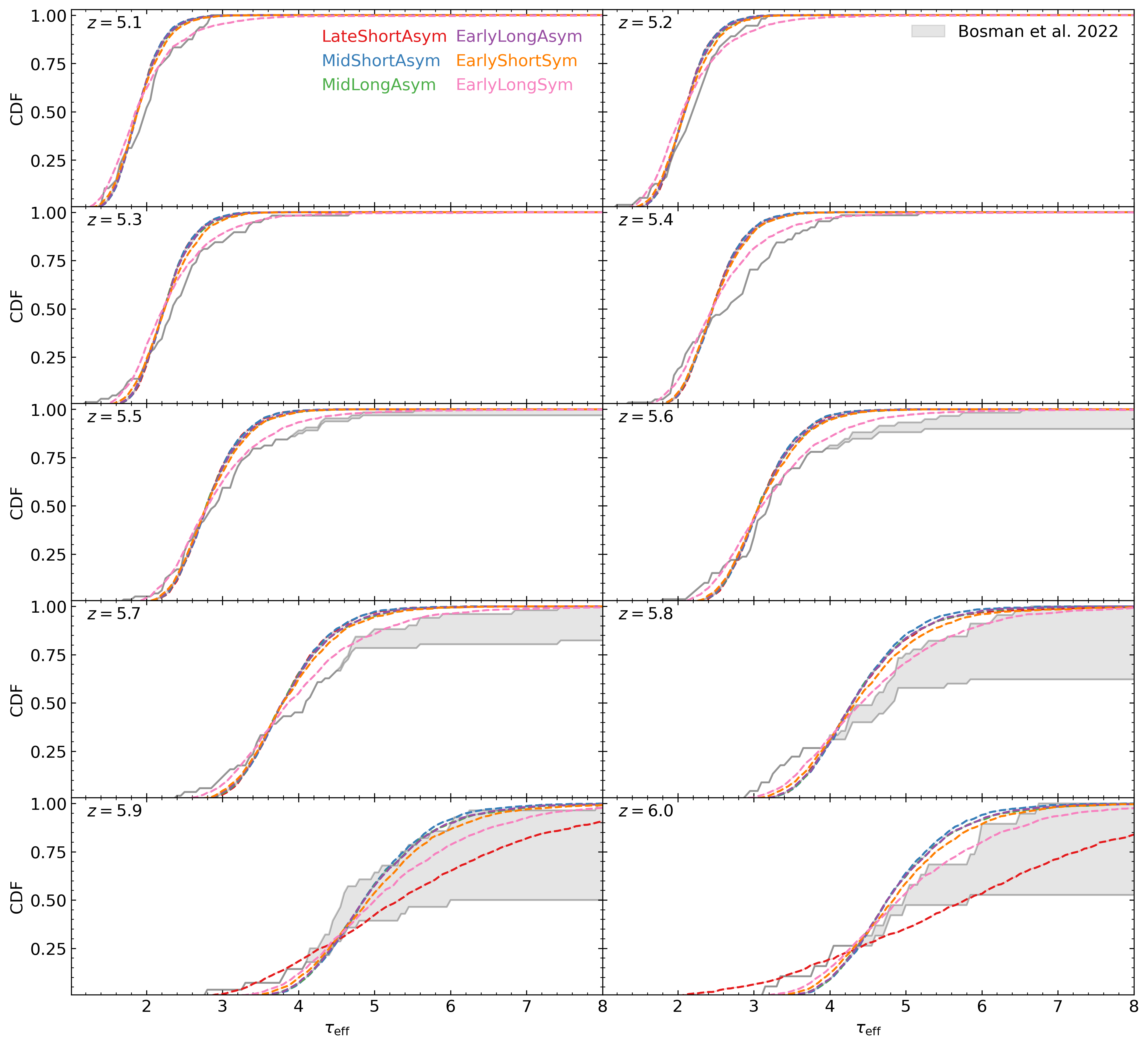}
\caption{Same as Figure~\ref{fig:tau_eff_cdf_hot}, but for the cold models. As with the autocorrelation functions, there are reduced inter-model differences in the case of cold models. Additionally, the cold models have narrower CDFs than the hot ones.}
\label{fig:tau_eff_cdf_cold}
\end{figure*}

We calculate the mean flux and the corresponding effective optical depth for each model skewer, and plot their cumulative distribution for the hot and cold models in Figures~\ref{fig:tau_eff_cdf_hot} and~\ref{fig:tau_eff_cdf_cold}. We overplot the measurements from~\citet{bosman22}, measured along 20 \mpchs segments of the XQR-30 sightlines. We adjust the ensemble mean flux to the average flux of all their sightlines, since this does not exactly match the reported mean flux per $\Delta z=0.1$ bin. For their data, we adhere to their convention for lower and upper limits, and these are indicated by the gray shading in the figures.\footnote{\citet{bosman22} limits are based on non-detections in their dataset, defined as lines of sight where the local mean flux is less than $2\sigma_{\langle F\rangle}$. For these non-detections, the bounds are determined by setting $\langle F \rangle_\mathrm{los}=0.0$ and $2\sigma_{\langle F\rangle}$.}

First considering the hot models, \textsl{MidAsym} and \textsl{EarlyAsym} are fairly similar to one another across the entire redshift range, differing mostly in the upper 40 percent of the CDF at the higher $\tau_\mathrm{eff}$ end of the distribution. Within this subset of models, the two \textsl{Long} ones tend to have a slightly wider distribution. The \textsl{Late} history is extremely wide at $5.9 < z < 6.0$, but has settled to match \textsl{MidShortAsym} by $z=5.8$, once it has finished reionizing and the Ly$\alpha$ opacity is no longer impacted by remaining opaque neutral regions. It is interesting that these two histories are identical in $A_z$ and $\Delta z$, and result in virtually identical CDFs despite their different $z_\mathrm{mid}$.

The \textsl{Sym} histories are consistently the most different from the other models, since they only complete reionization at $z<6$ they retain cells that are wholly opaque. \textsl{EarlyLongSym} is the most impacted by this, since it has not even reached one percent neutral by $z=5$. However, \textsl{EarlyShortSym} finishes reionizing at $z=5.4$, and remains a bit wider compared to the other models even by $z=5.0$. It seems that the longer the duration and the more uniform the distribution of reionization redshifts (center panel of Figure~\ref{fig:xHI_plot}), the wider the $\tau_\mathrm{eff}$ CDF, likely as a byproduct of the temperature fluctuations introduced by a prolonged reionization process. Of course, there is also an impact from the ongoing reionization in the \textsl{Sym} models, and this is why it will be necessary in the future to do a more thorough exploration of the reionization history parameter space.

The cold models in Figure~\ref{fig:tau_eff_cdf_cold} have narrower distributions than the hot ones, with the exception of the \textsl{Late} history. Here, it is apparent that the breadth of the $\tau_\mathrm{eff}$ CDF is related to the amplitude of the autocorrelation function, where a wider CDF is associated with a higher $\xi_F$, which is to be expected given that these are both probes of the opacity fluctuations.

The models approximately overlap~\citet{bosman22} best at $z\leq5.2$ and $z\geq5.8$, when the observed CDF is quite narrow and wide, respectively. In most of the other redshift bins, the models tend to be too narrow, missing sightlines with more extreme low and high opacities. The exception is the \textsl{EarlyLongSym} model, which is decently overlapping the upper limits, where the flux has been ``optimistically'' set to $2 \sigma_\mathrm{F}$, in all $z$ bins.

The excessive narrowness of the majority of the model CDFs, especially when combined with the systematically low $\xi_F$ values in the previous section, is indicative that in the future our models must introduce greater fluctuations in the temperature, neutral fraction, or UVB. These models intrinsically include fluctuations in temperature because of the inclusion of inhomogeneous reionization. However, given the trends in the $\tau_\mathrm{eff}$ CDF it seems likely that our use of highly asymmetric histories in the \textsl{Asym} models is suppressing temperature fluctuations, narrowing the distributions, causing a greater mismatch with~\citet{bosman22}. Additionally, while the \textsl{Sym} models approximately span the range of ``permitted'' reionization histories ending at $z<6$, there are intermediate histories that could also be considered. Lastly, these simulations do not contain a varying UVB, although this will be added in the future, and will certainly help to alleviate the discrepancy.

\section{Discussion}\label{sec:discussion}
In this work, we have used the semi-numerical code AMBER to implement inhomogeneous reionization histories into the cosmological hydrodynamics code \texttt{Nyx}. Running at a higher resolution ($\sim 10$ \kpch) than most comparable investigations, and using a variety of reionization timings, we have shown how certain histories can better reproduce various metrics of Ly$\alpha$ in the $5 < z < 6$ range, and how certain metrics are systematically discrepant regardless of the model.

It is common for numerical simulations to fail to reproduce the relatively high values of the $\log$\skm$>-0.9$ end of the Ly$\alpha$ flux power spectrum of~\citet{boera19}, and our hot models also reproduce this issue. However, we were able to overshoot the power on this scale by implementing histories with a lower heat injection than is commonly assumed while still decently reproducing the observationally-derived $T_0$. Thus it seems that some intermediate temperature could be found to match the power precisely. However, these results only further confirm those of previous studies, to emphasize the degeneracies inherent in these metrics~\citep[e.g.][]{wu19}, since despite the wildly different pressure smoothing scales in our hot models, their $z=5.0$ power spectra are essentially indistinguishable. This motivates the need for experiments involving careful inference over many reionization models covering a wider range of the parameters considered here.

Related to this,~\citet{khan23} recently investigated the effect of initial conditions on Ly$\alpha$ forest power and found that use of a glass particle initialization scheme can produce a large excess in the power at $\log$\skm$>-1.3$, up to 50 percent at higher $z$. This appears to result from the impact of particle coupling on both the resultant density field and peculiar velocities, and in an SPH simulation may be reduced by using the gravitational softening length of SPH particles (adaptive softening) rather than using the scale determined by dark matter particles. We use glass initialization for both dark matter and gas particles, with the gas particles having a staggered displacement with respect to the dark matter. Although our setup is significantly different than in~\citet{khan23}, it is possible that our high $k$ power could arise from this unphysical coupling effect. If this is the case, then our cold models with their currently excessive high $k$ power would likely provide a decent match to the~\citet{boera19} $z=5$ power spectrum, disfavoring the commonly assumed $\Delta T_\mathrm{re}$ of 20000 K.

The temperature at mean density is not directly observable, but still provides a useful point of comparison for the models, especially since the amount of heat injection is not certain. The majority of observations find quite low $T_0$ from $5.0 < z < 5.8$, less than 9000 K, which several of our models match at $z\sim5.0$ but overshoot at higher redshifts. These have varying heat injection and reionization histories. The hotter derived $T_0$ values of~\citet{gaikwad20} are matched by all hot models except for the very late and rapid \textsl{LateShortAsym} history, which is far \emph{too} hot. It doesn't seem possible to achieve the very low temperatures of~\citet{garzilli17} or~\citet{walther19}, unless using a very low heat injection, leading to a flat evolution at $5 < z < 6$. The~\citet{gaikwad20} measurements are often interpreted as suggesting a decrease in temperature and thus a period of significant cooling following the end of reionization, which is only a feature of the hot histories, and is also stronger for \textsl{Asym} histories. Generally though, it seems impossible to choose a preferred reionization history on the basis of the observationally-derived temperature at mean density, until they begin to converge to some distinctive trend.

Considering the more direct measures of gas phase, Ly$\alpha$ opacity metrics, we find that the metrics appear to be sensitive to different details of the reionization history. Histories where the reionization midpoint occurs later have less relaxed $\rho-T$ distributions and smaller pressure smoothing scales, which leaves the potential for them to have increased small scale fluctuations. As a result, we can see that the small scale $P_\mathrm{Ly\alpha}$ in $\Delta T_\mathrm{re}=10,000$ K reionization histories can give an indication of the reionization midpoint, and the pressure smoothing scale. However, in practice these fluctuations are only observable if $\Delta T_\mathrm{re}$ is sufficiently low, less than $20,000$ K, otherwise they are completely washed out in the 1D flux power spectrum. $\xi_F$ and the $\tau_\mathrm{eff}$ CDF on the other hand appear to be sensitive to the duration $\Delta z$ and asymmetry $A_z$ of the histories, with a longer or more symmetric history leading to higher $\xi_F$ and wider CDFs. This may be because the more varied timing in heat injections leads to greater temperature fluctuations.

Our simulations can span the 1D flux power spectrum measurements easily, but we are less able to match the autocorrelation functions and effective optical depth distributions. Given the trends we see with regard to $\Delta z$ and $A_z$, exploration of more \textsl{Sym}-like histories could partially alleviate this. However, the main issue is likely the lack of UVB fluctuations in this model, as~\citet{wolfson23a} demonstrated that more fluctuations through a smaller mean free path would increase the autocorrelation function values across the range of lags considered here. Further, this change would primarily impact large scales~\citep{onorbe19}, thus minimally affecting the $k$ modes covered in our 1D flux power spectra. This supports results from both~\citet{bosman22} and~\citet{zhu22} that rule out a homogeneous UVB for $z\gtrapprox 5.4$.

Most of our reionization histories were chosen on the basis that they adhered to the \emph{Planck} FlexKnot~\citep[originating in][]{millea18} limits in volume-weighted neutral fraction, which only permits ``knots'' at $z>6$ in order to match the reionization redshift range implied by observations of Gunn-Peterson troughs. It thus enforces $x_\mathrm{HI,v}<0.02$ by $z=5.8$, which is arguably not the best assumption given the current Ly$\alpha$ forest evidence for a late reionization. We consider only one model that violates this result, and instead adheres to the dark pixel fraction measurements from~\citet{mcgreer15} and the directly measured $\tau_\mathrm{e}$.
In the future, a better choice would be to explore a larger number of \textsl{EarlyLongSym}-like models, that permit $x_\mathrm{HI,v}>0$ for $z<6$.

\section{Conclusions}\label{sec:conclusions}
In this work we have evaluated how a variety of reionization histories affect the IGM in \texttt{Nyx} cosmological simulations and several resultant Ly$\alpha$ forest metrics: the 1D flux power spectrum, autocorrelation function, and effective optical depth distribution. We accomplished this using the semi-numerical code AMBER, which allowed us to parametrize the evolution of the neutral fraction with redshift with the shape of the reionization histories rather than the details of the physics. We have found that:

\begin{itemize}

\item Reionization histories with longer durations and greater symmetry produce gas that spans a wide range of properties in $\rho-T$ space. This naturally will lead to more fluctuations in the temperature and density fields.

\item The ensemble pressure smoothing scale measured in the IGM grows linearly with time since the mass-weighted midpoint, and hotter heat injection (20,000 versus 10,000 K) leads to a faster growth. This scale is also correlated with the adiabatic index $\gamma$ as measured from the $\rho-T$ distribution. Low heat injection models also have a higher adiabatic index for a given pressure smoothing scale.

\item Models with lower heat injection at reionization (10,000 K) have shorter pressure smoothing scales in the IGM, and their average temperature post-reionization is low enough that the remaining small scale power is still visible in the 1D Ly$\alpha$ flux power spectrum. 

\item Hot ($\Delta T_\mathrm{re}=20000$ K) reionization histories better reproduce the shape of the observed 1D Ly$\alpha$ flux power spectrum than cold ($\Delta T_\mathrm{re}=10000$ K) histories. However, hot histories also underproduce observed $\log$\skm$>-0.9$ power, while the cold ones overproduce it.

\item All of the models systematically underestimate the observed autocorrelation function, and the disagreement is worse at $5.3 \leq z \leq 5.8$ than at $z\leq5.2$ and $z\geq 5.9$. The models that tend to match the best are those that don't finish reionizing until $z\leq5.4$ and that have more symmetric reionization histories, or an approximately normal distribution of reionization redshifts. Hotter $\Delta T_\mathrm{re}$ models produce higher autocorrelation values than cold ones, but only at $\Delta v<500$ km/s.

\item Similar to the autocorrelation function, the models underpredict the observed width of the distribution of effective optical depth measurements, with the exception of our extreme history, that does not even reach $x_\mathrm{HI,v}=1$ per cent by $z=5.0$. Simulations with similar durations and asymmetries produce similar $\tau_\mathrm{eff}$ CDFs, suggesting that the differences here arise from temperature or density fluctuations in the post-reionization IGM.

\end{itemize}

Notably, our experimental setup includes fluctuations in temperature and hydrogen neutral fraction, but excludes any kind of ionizing background fluctuation. Greater spatial variation in the UVB is known to contribute to a higher $\xi_F$ and wider CDFs, and has already been deemed necessary for matching observations down to at least $z=5.3$~\citep{bosman22,zhu22}. Thus, in order to properly compare to observations, it seems absolutely necessary to include UVB fluctuations.

This work demonstrates the sensitivity of the Ly$\alpha$ forest metrics to the reionization history, but also shows the situations where the effects of model parameters become degenerate. In order to place quantitative constraints on the history, it will be necessary to use a larger model grid that captures all of the competing physical effects, and with fewer arbitrary restrictions placed on the histories, such as forcing an end at $z\geq6$. 

\section*{Acknowledgements}
CCD thanks Hyunbae Park and Jean Sexton for assistance in implementing inhomogeneous reionizations in \texttt{Nyx}. CCD thanks Hy Trac for assistance in working with AMBER, and Sarah Bosman for providing re-calculated effective optical depth distributions. This research used resources of the National Energy Research Scientific Computing Center, which is supported by the Office of Science of the U.S. Department of Energy under Contract No. DE-AC02-05CH11231. This research further used resources of the Oak Ridge Leadership Computing Facility at the Oak Ridge National Laboratory, which is supported by the Office of Science of the U.S. Department of Energy under Contract No. DE-AC05-00OR22725.

\section*{Data Availability}
The data will be shared on reasonable request to the corresponding author.



\bibliographystyle{mnras}
\bibliography{inhomogeneous_reion_obs_comparison} 

\begin{thebibliography}{}
\makeatletter
\relax
\def\mn@urlcharsother{\let\do\@makeother \do\$\do\&\do\#\do\^\do\_\do\%\do\~}
\def\mn@doi{\begingroup\mn@urlcharsother \@ifnextchar [ {\mn@doi@} {\mn@doi@[]}}
\def\mn@doi@[#1]#2{\def\@tempa{#1}\ifx\@tempa\@empty \href {http://dx.doi.org/#2} {doi:#2}\else \href {http://dx.doi.org/#2} {#1}\fi \endgroup}
\def\mn@eprint#1#2{\mn@eprint@#1:#2::\@nil}
\def\mn@eprint@arXiv#1{\href {http://arxiv.org/abs/#1} {{\tt arXiv:#1}}}
\def\mn@eprint@dblp#1{\href {http://dblp.uni-trier.de/rec/bibtex/#1.xml} {dblp:#1}}
\def\mn@eprint@#1:#2:#3:#4\@nil{\def\@tempa {#1}\def\@tempb {#2}\def\@tempc {#3}\ifx \@tempc \@empty \let \@tempc \@tempb \let \@tempb \@tempa \fi \ifx \@tempb \@empty \def\@tempb {arXiv}\fi \@ifundefined {mn@eprint@\@tempb}{\@tempb:\@tempc}{\expandafter \expandafter \csname mn@eprint@\@tempb\endcsname \expandafter{\@tempc}}}

\bibitem[\protect\citeauthoryear{{Almgren}, {Bell}, {Lijewski}, {Luki{\'c}}  \& {Van Andel}}{{Almgren} et~al.}{2013}]{almgren13}
{Almgren} A.~S.,  {Bell} J.~B.,  {Lijewski} M.~J.,  {Luki{\'c}} Z.,   {Van Andel} E.,  2013, \mn@doi [\apj] {10.1088/0004-637X/765/1/39}, \href {https://ui.adsabs.harvard.edu/abs/2013ApJ...765...39A} {765, 39}

\bibitem[\protect\citeauthoryear{{Ba{\~n}ados} et~al.,}{{Ba{\~n}ados} et~al.}{2018}]{banados18}
{Ba{\~n}ados} E.,  et~al., 2018, \mn@doi [\nat] {10.1038/nature25180}, \href {https://ui.adsabs.harvard.edu/abs/2018Natur.553..473B} {553, 473}

\bibitem[\protect\citeauthoryear{{Battaglia}, {Trac}, {Cen}  \& {Loeb}}{{Battaglia} et~al.}{2013}]{battaglia13}
{Battaglia} N.,  {Trac} H.,  {Cen} R.,   {Loeb} A.,  2013, \mn@doi [\apj] {10.1088/0004-637X/776/2/81}, \href {https://ui.adsabs.harvard.edu/abs/2013ApJ...776...81B} {776, 81}

\bibitem[\protect\citeauthoryear{{Becker}, {D'Aloisio}, {Christenson}, {Zhu}, {Worseck}  \& {Bolton}}{{Becker} et~al.}{2021}]{becker21}
{Becker} G.~D.,  {D'Aloisio} A.,  {Christenson} H.~M.,  {Zhu} Y.,  {Worseck} G.,   {Bolton} J.~S.,  2021, \mn@doi [\mnras] {10.1093/mnras/stab2696}, \href {https://ui.adsabs.harvard.edu/abs/2021MNRAS.508.1853B} {508, 1853}

\bibitem[\protect\citeauthoryear{{Boera}, {Becker}, {Bolton}  \& {Nasir}}{{Boera} et~al.}{2019}]{boera19}
{Boera} E.,  {Becker} G.~D.,  {Bolton} J.~S.,   {Nasir} F.,  2019, \mn@doi [\apj] {10.3847/1538-4357/aafee4}, \href {https://ui.adsabs.harvard.edu/abs/2019ApJ...872..101B} {872, 101}

\bibitem[\protect\citeauthoryear{{Bosman} et~al.,}{{Bosman} et~al.}{2022}]{bosman22}
{Bosman} S. E.~I.,  et~al., 2022, \mn@doi [\mnras] {10.1093/mnras/stac1046}, \href {https://ui.adsabs.harvard.edu/abs/2022MNRAS.514...55B} {514, 55}

\bibitem[\protect\citeauthoryear{{Bouchet}, {Colombi}, {Hivon}  \& {Juszkiewicz}}{{Bouchet} et~al.}{1995}]{bouchet95}
{Bouchet} F.~R.,  {Colombi} S.,  {Hivon} E.,   {Juszkiewicz} R.,  1995, \mn@doi [\aap] {10.48550/arXiv.astro-ph/9406013}, \href {https://ui.adsabs.harvard.edu/abs/1995A&A...296..575B} {296, 575}

\bibitem[\protect\citeauthoryear{{Bruton}, {Lin}, {Scarlata}  \& {Hayes}}{{Bruton} et~al.}{2023}]{bruton23}
{Bruton} S.,  {Lin} Y.-H.,  {Scarlata} C.,   {Hayes} M.~J.,  2023, \mn@doi [arXiv e-prints] {10.48550/arXiv.2303.03419}, \href {https://ui.adsabs.harvard.edu/abs/2023arXiv230303419B} {p. arXiv:2303.03419}

\bibitem[\protect\citeauthoryear{{Chabanier} et~al.,}{{Chabanier} et~al.}{2023}]{chabanier23}
{Chabanier} S.,  et~al., 2023, \mn@doi [\mnras] {10.1093/mnras/stac3294}, \href {https://ui.adsabs.harvard.edu/abs/2023MNRAS.518.3754C} {518, 3754}

\bibitem[\protect\citeauthoryear{{Costa}, {Sijacki}, {Trenti}  \& {Haehnelt}}{{Costa} et~al.}{2014}]{costa14}
{Costa} T.,  {Sijacki} D.,  {Trenti} M.,   {Haehnelt} M.~G.,  2014, \mn@doi [\mnras] {10.1093/mnras/stu101}, \href {https://ui.adsabs.harvard.edu/abs/2014MNRAS.439.2146C} {439, 2146}

\bibitem[\protect\citeauthoryear{{Davies} et~al.,}{{Davies} et~al.}{2018}]{davies18}
{Davies} F.~B.,  et~al., 2018, \mn@doi [\apj] {10.3847/1538-4357/aad6dc}, \href {https://ui.adsabs.harvard.edu/abs/2018ApJ...864..142D} {864, 142}

\bibitem[\protect\citeauthoryear{{DeBoer} et~al.,}{{DeBoer} et~al.}{2017}]{deboer17}
{DeBoer} D.~R.,  et~al., 2017, \mn@doi [\pasp] {10.1088/1538-3873/129/974/045001}, \href {https://ui.adsabs.harvard.edu/abs/2017PASP..129d5001D} {129, 045001}

\bibitem[\protect\citeauthoryear{{Doughty}, {Hennawi}, {Davies}, {Luki{\'c}}  \& {O{\~n}orbe}}{{Doughty} et~al.}{2023}]{doughty23}
{Doughty} C.~C.,  {Hennawi} J.~F.,  {Davies} F.~B.,  {Luki{\'c}} Z.,   {O{\~n}orbe} J.,  2023, \mn@doi [arXiv e-prints] {10.48550/arXiv.2305.16200}, \href {https://ui.adsabs.harvard.edu/abs/2023arXiv230516200D} {p. arXiv:2305.16200}

\bibitem[\protect\citeauthoryear{{Fan}, {Ba{\~n}ados}  \& {Simcoe}}{{Fan} et~al.}{2023}]{fan23}
{Fan} X.,  {Ba{\~n}ados} E.,   {Simcoe} R.~A.,  2023, \mn@doi [\araa] {10.1146/annurev-astro-052920-102455}, \href {https://ui.adsabs.harvard.edu/abs/2023ARA&A..61..373F} {61, 373}

\bibitem[\protect\citeauthoryear{{Feng}, {Chu}, {Seljak}  \& {McDonald}}{{Feng} et~al.}{2016}]{feng16}
{Feng} Y.,  {Chu} M.-Y.,  {Seljak} U.,   {McDonald} P.,  2016, \mn@doi [\mnras] {10.1093/mnras/stw2123}, \href {https://ui.adsabs.harvard.edu/abs/2016MNRAS.463.2273F} {463, 2273}

\bibitem[\protect\citeauthoryear{{Gaikwad} et~al.,}{{Gaikwad} et~al.}{2020}]{gaikwad20}
{Gaikwad} P.,  et~al., 2020, \mn@doi [\mnras] {10.1093/mnras/staa907}, \href {https://ui.adsabs.harvard.edu/abs/2020MNRAS.494.5091G} {494, 5091}

\bibitem[\protect\citeauthoryear{{Gaikwad} et~al.,}{{Gaikwad} et~al.}{2023}]{gaikwad23}
{Gaikwad} P.,  et~al., 2023, \mn@doi [arXiv e-prints] {10.48550/arXiv.2304.02038}, \href {https://ui.adsabs.harvard.edu/abs/2023arXiv230402038G} {p. arXiv:2304.02038}

\bibitem[\protect\citeauthoryear{{Garzilli}, {Boyarsky}  \& {Ruchayskiy}}{{Garzilli} et~al.}{2017}]{garzilli17}
{Garzilli} A.,  {Boyarsky} A.,   {Ruchayskiy} O.,  2017, \mn@doi [Physics Letters B] {10.1016/j.physletb.2017.08.022}, \href {https://ui.adsabs.harvard.edu/abs/2017PhLB..773..258G} {773, 258}

\bibitem[\protect\citeauthoryear{{Gnedin} \& {Hui}}{{Gnedin} \& {Hui}}{1998}]{gnedin98}
{Gnedin} N.~Y.,  {Hui} L.,  1998, \mn@doi [\mnras] {10.1046/j.1365-8711.1998.01249.x}, \href {https://ui.adsabs.harvard.edu/abs/1998MNRAS.296...44G} {296, 44}

\bibitem[\protect\citeauthoryear{{Greig}, {Mesinger}, {Haiman}  \& {Simcoe}}{{Greig} et~al.}{2017}]{greig17}
{Greig} B.,  {Mesinger} A.,  {Haiman} Z.,   {Simcoe} R.~A.,  2017, \mn@doi [\mnras] {10.1093/mnras/stw3351}, \href {https://ui.adsabs.harvard.edu/abs/2017MNRAS.466.4239G} {466, 4239}

\bibitem[\protect\citeauthoryear{{Greig}, {Mesinger}  \& {Ba{\~n}ados}}{{Greig} et~al.}{2019}]{greig19}
{Greig} B.,  {Mesinger} A.,   {Ba{\~n}ados} E.,  2019, \mn@doi [\mnras] {10.1093/mnras/stz230}, \href {https://ui.adsabs.harvard.edu/abs/2019MNRAS.484.5094G} {484, 5094}

\bibitem[\protect\citeauthoryear{{Greig}, {Mesinger}, {Davies}, {Wang}, {Yang}  \& {Hennawi}}{{Greig} et~al.}{2022}]{greig22}
{Greig} B.,  {Mesinger} A.,  {Davies} F.~B.,  {Wang} F.,  {Yang} J.,   {Hennawi} J.~F.,  2022, \mn@doi [\mnras] {10.1093/mnras/stac825}, \href {https://ui.adsabs.harvard.edu/abs/2022MNRAS.512.5390G} {512, 5390}

\bibitem[\protect\citeauthoryear{{HERA Collaboration} et~al.,}{{HERA Collaboration} et~al.}{2023}]{hera23}
{HERA Collaboration} et~al., 2023, \mn@doi [\apj] {10.3847/1538-4357/acaf50}, \href {https://ui.adsabs.harvard.edu/abs/2023ApJ...945..124H} {945, 124}

\bibitem[\protect\citeauthoryear{{Heintz} et~al.,}{{Heintz} et~al.}{2023}]{heintz23}
{Heintz} K.~E.,  et~al., 2023, \mn@doi [arXiv e-prints] {10.48550/arXiv.2306.00647}, \href {https://ui.adsabs.harvard.edu/abs/2023arXiv230600647H} {p. arXiv:2306.00647}

\bibitem[\protect\citeauthoryear{{Hoag} et~al.,}{{Hoag} et~al.}{2019}]{hoag19}
{Hoag} A.,  et~al., 2019, \mn@doi [\apj] {10.3847/1538-4357/ab1de7}, \href {https://ui.adsabs.harvard.edu/abs/2019ApJ...878...12H} {878, 12}

\bibitem[\protect\citeauthoryear{{Hockney} \& {Eastwood}}{{Hockney} \& {Eastwood}}{1988}]{hockney88}
{Hockney} R.~W.,  {Eastwood} J.~W.,  1988, {Computer simulation using particles}

\bibitem[\protect\citeauthoryear{{Hui} \& {Gnedin}}{{Hui} \& {Gnedin}}{1997}]{hui97}
{Hui} L.,  {Gnedin} N.~Y.,  1997, \mn@doi [\mnras] {10.1093/mnras/292.1.27}, \href {https://ui.adsabs.harvard.edu/abs/1997MNRAS.292...27H} {292, 27}

\bibitem[\protect\citeauthoryear{{Jacobus}, {Harrington}  \& {Luki{\'c}}}{{Jacobus} et~al.}{2023}]{jacobus23}
{Jacobus} C.,  {Harrington} P.,   {Luki{\'c}} Z.,  2023, \mn@doi [arXiv e-prints] {10.48550/arXiv.2308.02637}, \href {https://ui.adsabs.harvard.edu/abs/2023arXiv230802637J} {p. arXiv:2308.02637}

\bibitem[\protect\citeauthoryear{{Jin} et~al.,}{{Jin} et~al.}{2023}]{jin23}
{Jin} X.,  et~al., 2023, \mn@doi [\apj] {10.3847/1538-4357/aca678}, \href {https://ui.adsabs.harvard.edu/abs/2023ApJ...942...59J} {942, 59}

\bibitem[\protect\citeauthoryear{{Kannan}, {Garaldi}, {Smith}, {Pakmor}, {Springel}, {Vogelsberger}  \& {Hernquist}}{{Kannan} et~al.}{2022}]{kannan22}
{Kannan} R.,  {Garaldi} E.,  {Smith} A.,  {Pakmor} R.,  {Springel} V.,  {Vogelsberger} M.,   {Hernquist} L.,  2022, \mn@doi [\mnras] {10.1093/mnras/stab3710}, \href {https://ui.adsabs.harvard.edu/abs/2022MNRAS.511.4005K} {511, 4005}

\bibitem[\protect\citeauthoryear{{Kara{\c{c}}ayl{\i}} et~al.,}{{Kara{\c{c}}ayl{\i}} et~al.}{2022}]{karacayli22}
{Kara{\c{c}}ayl{\i}} N.~G.,  et~al., 2022, \mn@doi [\mnras] {10.1093/mnras/stab3201}, \href {https://ui.adsabs.harvard.edu/abs/2022MNRAS.509.2842K} {509, 2842}

\bibitem[\protect\citeauthoryear{{Keating}, {Bolton}, {Cullen}, {Haehnelt}, {Puchwein}  \& {Kulkarni}}{{Keating} et~al.}{2023}]{keating23}
{Keating} L.~C.,  {Bolton} J.~S.,  {Cullen} F.,  {Haehnelt} M.~G.,  {Puchwein} E.,   {Kulkarni} G.,  2023, \mn@doi [arXiv e-prints] {10.48550/arXiv.2308.05800}, \href {https://ui.adsabs.harvard.edu/abs/2023arXiv230805800K} {p. arXiv:2308.05800}

\bibitem[\protect\citeauthoryear{{Khan}, {Kulkarni}, {Bolton}, {Haehnelt}, {Ir{\v{s}}i{\v{c}}}, {Puchwein}  \& {Asthana}}{{Khan} et~al.}{2023}]{khan23}
{Khan} N.~K.,  {Kulkarni} G.,  {Bolton} J.~S.,  {Haehnelt} M.~G.,  {Ir{\v{s}}i{\v{c}}} V.,  {Puchwein} E.,   {Asthana} S.,  2023, \mn@doi [arXiv e-prints] {10.48550/arXiv.2310.07767}, \href {https://ui.adsabs.harvard.edu/abs/2023arXiv231007767K} {p. arXiv:2310.07767}

\bibitem[\protect\citeauthoryear{{Kulkarni}, {Hennawi}, {O{\~n}orbe}, {Rorai}  \& {Springel}}{{Kulkarni} et~al.}{2015}]{kulkarni15}
{Kulkarni} G.,  {Hennawi} J.~F.,  {O{\~n}orbe} J.,  {Rorai} A.,   {Springel} V.,  2015, \mn@doi [\apj] {10.1088/0004-637X/812/1/30}, \href {https://ui.adsabs.harvard.edu/abs/2015ApJ...812...30K} {812, 30}

\bibitem[\protect\citeauthoryear{{Lewis}, {Challinor}  \& {Lasenby}}{{Lewis} et~al.}{2000}]{lewis00}
{Lewis} A.,  {Challinor} A.,   {Lasenby} A.,  2000, \mn@doi [\apj] {10.1086/309179}, \href {https://ui.adsabs.harvard.edu/abs/2000ApJ...538..473L} {538, 473}

\bibitem[\protect\citeauthoryear{{Loeb} \& {Barkana}}{{Loeb} \& {Barkana}}{2001}]{loeb01}
{Loeb} A.,  {Barkana} R.,  2001, \mn@doi [\araa] {10.1146/annurev.astro.39.1.19}, \href {https://ui.adsabs.harvard.edu/abs/2001ARA&A..39...19L} {39, 19}

\bibitem[\protect\citeauthoryear{{Luki{\'c}}, {Stark}, {Nugent}, {White}, {Meiksin}  \& {Almgren}}{{Luki{\'c}} et~al.}{2015}]{lukic15}
{Luki{\'c}} Z.,  {Stark} C.~W.,  {Nugent} P.,  {White} M.,  {Meiksin} A.~A.,   {Almgren} A.,  2015, \mn@doi [\mnras] {10.1093/mnras/stu2377}, \href {https://ui.adsabs.harvard.edu/abs/2015MNRAS.446.3697L} {446, 3697}

\bibitem[\protect\citeauthoryear{{Mason}, {Treu}, {Dijkstra}, {Mesinger}, {Trenti}, {Pentericci}, {de Barros}  \& {Vanzella}}{{Mason} et~al.}{2018}]{mason18}
{Mason} C.~A.,  {Treu} T.,  {Dijkstra} M.,  {Mesinger} A.,  {Trenti} M.,  {Pentericci} L.,  {de Barros} S.,   {Vanzella} E.,  2018, \mn@doi [\apj] {10.3847/1538-4357/aab0a7}, \href {https://ui.adsabs.harvard.edu/abs/2018ApJ...856....2M} {856, 2}

\bibitem[\protect\citeauthoryear{{McGreer}, {Mesinger}  \& {D'Odorico}}{{McGreer} et~al.}{2015}]{mcgreer15}
{McGreer} I.~D.,  {Mesinger} A.,   {D'Odorico} V.,  2015, \mn@doi [\mnras] {10.1093/mnras/stu2449}, \href {https://ui.adsabs.harvard.edu/abs/2015MNRAS.447..499M} {447, 499}

\bibitem[\protect\citeauthoryear{{Mesinger}, {Furlanetto}  \& {Cen}}{{Mesinger} et~al.}{2011}]{mesinger11}
{Mesinger} A.,  {Furlanetto} S.,   {Cen} R.,  2011, \mn@doi [\mnras] {10.1111/j.1365-2966.2010.17731.x}, \href {https://ui.adsabs.harvard.edu/abs/2011MNRAS.411..955M} {411, 955}

\bibitem[\protect\citeauthoryear{{Millea} \& {Bouchet}}{{Millea} \& {Bouchet}}{2018}]{millea18}
{Millea} M.,  {Bouchet} F.,  2018, \mn@doi [\aap] {10.1051/0004-6361/201833288}, \href {https://ui.adsabs.harvard.edu/abs/2018A&A...617A..96M} {617, A96}

\bibitem[\protect\citeauthoryear{{O{\~n}orbe}, {Davies}, {Luki{\'c}}, {}, {Hennawi}  \& {Sorini}}{{O{\~n}orbe} et~al.}{2019}]{onorbe19}
{O{\~n}orbe} J.,  {Davies} F.~B.,  {Luki{\'c}} {} Z.,  {Hennawi} J.~F.,   {Sorini} D.,  2019, \mn@doi [\mnras] {10.1093/mnras/stz984}, \href {https://ui.adsabs.harvard.edu/abs/2019MNRAS.486.4075O} {486, 4075}

\bibitem[\protect\citeauthoryear{{O'Leary} \& {McQuinn}}{{O'Leary} \& {McQuinn}}{2012}]{oleary12}
{O'Leary} R.~M.,  {McQuinn} M.,  2012, \mn@doi [\apj] {10.1088/0004-637X/760/1/4}, \href {https://ui.adsabs.harvard.edu/abs/2012ApJ...760....4O} {760, 4}

\bibitem[\protect\citeauthoryear{{Ouchi} et~al.,}{{Ouchi} et~al.}{2010}]{ouchi10}
{Ouchi} M.,  et~al., 2010, \mn@doi [\apj] {10.1088/0004-637X/723/1/869}, \href {https://ui.adsabs.harvard.edu/abs/2010ApJ...723..869O} {723, 869}

\bibitem[\protect\citeauthoryear{{Planck Collaboration} et~al.,}{{Planck Collaboration} et~al.}{2020}]{planck20}
{Planck Collaboration} et~al., 2020, \mn@doi [\aap] {10.1051/0004-6361/201833910}, \href {https://ui.adsabs.harvard.edu/abs/2020A&A...641A...6P} {641, A6}

\bibitem[\protect\citeauthoryear{{Puchwein} et~al.,}{{Puchwein} et~al.}{2023}]{puchwein23}
{Puchwein} E.,  et~al., 2023, \mn@doi [\mnras] {10.1093/mnras/stac3761}, \href {https://ui.adsabs.harvard.edu/abs/2023MNRAS.519.6162P} {519, 6162}

\bibitem[\protect\citeauthoryear{{Rosdahl} et~al.,}{{Rosdahl} et~al.}{2018}]{rosdahl18}
{Rosdahl} J.,  et~al., 2018, \mn@doi [\mnras] {10.1093/mnras/sty1655}, \href {https://ui.adsabs.harvard.edu/abs/2018MNRAS.479..994R} {479, 994}

\bibitem[\protect\citeauthoryear{{Rosdahl} et~al.,}{{Rosdahl} et~al.}{2022}]{rosdahl22}
{Rosdahl} J.,  et~al., 2022, \mn@doi [\mnras] {10.1093/mnras/stac1942}, \href {https://ui.adsabs.harvard.edu/abs/2022MNRAS.515.2386R} {515, 2386}

\bibitem[\protect\citeauthoryear{{Scoccimarro}}{{Scoccimarro}}{1998}]{scoccimarro98}
{Scoccimarro} R.,  1998, \mn@doi [\mnras] {10.1046/j.1365-8711.1998.01845.x}, \href {https://ui.adsabs.harvard.edu/abs/1998MNRAS.299.1097S} {299, 1097}

\bibitem[\protect\citeauthoryear{{Sexton}, {Lukic}, {Almgren}, {Daley}, {Friesen}, {Myers}  \& {Zhang}}{{Sexton} et~al.}{2021}]{sexton21}
{Sexton} J.,  {Lukic} Z.,  {Almgren} A.,  {Daley} C.,  {Friesen} B.,  {Myers} A.,   {Zhang} W.,  2021, \mn@doi [The Journal of Open Source Software] {10.21105/joss.03068}, \href {https://ui.adsabs.harvard.edu/abs/2021JOSS....6.3068S} {6, 3068}

\bibitem[\protect\citeauthoryear{{Trac}, {Chen}, {Holst}, {Alvarez}  \& {Cen}}{{Trac} et~al.}{2022}]{trac22}
{Trac} H.,  {Chen} N.,  {Holst} I.,  {Alvarez} M.~A.,   {Cen} R.,  2022, \mn@doi [\apj] {10.3847/1538-4357/ac5116}, \href {https://ui.adsabs.harvard.edu/abs/2022ApJ...927..186T} {927, 186}

\bibitem[\protect\citeauthoryear{{Viel}, {Becker}, {Bolton}  \& {Haehnelt}}{{Viel} et~al.}{2013a}]{viel13b}
{Viel} M.,  {Becker} G.~D.,  {Bolton} J.~S.,   {Haehnelt} M.~G.,  2013a, \mn@doi [\prd] {10.1103/PhysRevD.88.043502}, \href {https://ui.adsabs.harvard.edu/abs/2013PhRvD..88d3502V} {88, 043502}

\bibitem[\protect\citeauthoryear{{Viel}, {Schaye}  \& {Booth}}{{Viel} et~al.}{2013b}]{viel13a}
{Viel} M.,  {Schaye} J.,   {Booth} C.~M.,  2013b, \mn@doi [\mnras] {10.1093/mnras/sts465}, \href {https://ui.adsabs.harvard.edu/abs/2013MNRAS.429.1734V} {429, 1734}

\bibitem[\protect\citeauthoryear{{Walther}, {Hennawi}, {Hiss}, {O{\~n}orbe}, {Lee}, {Rorai}  \& {O'Meara}}{{Walther} et~al.}{2018}]{walther18}
{Walther} M.,  {Hennawi} J.~F.,  {Hiss} H.,  {O{\~n}orbe} J.,  {Lee} K.-G.,  {Rorai} A.,   {O'Meara} J.,  2018, \mn@doi [\apj] {10.3847/1538-4357/aa9c81}, \href {https://ui.adsabs.harvard.edu/abs/2018ApJ...852...22W} {852, 22}

\bibitem[\protect\citeauthoryear{{Walther}, {O{\~n}orbe}, {Hennawi}  \& {Luki{\'c}}}{{Walther} et~al.}{2019}]{walther19}
{Walther} M.,  {O{\~n}orbe} J.,  {Hennawi} J.~F.,   {Luki{\'c}} Z.,  2019, \mn@doi [\apj] {10.3847/1538-4357/aafad1}, \href {https://ui.adsabs.harvard.edu/abs/2019ApJ...872...13W} {872, 13}

\bibitem[\protect\citeauthoryear{{Wang} et~al.,}{{Wang} et~al.}{2020}]{wang20}
{Wang} F.,  et~al., 2020, \mn@doi [\apj] {10.3847/1538-4357/ab8c45}, \href {https://ui.adsabs.harvard.edu/abs/2020ApJ...896...23W} {896, 23}

\bibitem[\protect\citeauthoryear{{Weibull}}{{Weibull}}{1951}]{weibull51}
{Weibull} W.,  1951, \mn@doi [Journal of Applied Mechanics] {10.1115/1.4010337}, \href {https://ui.adsabs.harvard.edu/abs/1951JAM....18..293W} {18, 293}

\bibitem[\protect\citeauthoryear{{Wolfson} et~al.,}{{Wolfson} et~al.}{2023a}]{wolfson23b}
{Wolfson} M.,  et~al., 2023a, \mn@doi [arXiv e-prints] {10.48550/arXiv.2309.03341}, \href {https://ui.adsabs.harvard.edu/abs/2023arXiv230903341W} {p. arXiv:2309.03341}

\bibitem[\protect\citeauthoryear{{Wolfson}, {Hennawi}, {Davies}  \& {O{\~n}orbe}}{{Wolfson} et~al.}{2023b}]{wolfson23a}
{Wolfson} M.,  {Hennawi} J.~F.,  {Davies} F.~B.,   {O{\~n}orbe} J.,  2023b, \mn@doi [\mnras] {10.1093/mnras/stad701}, \href {https://ui.adsabs.harvard.edu/abs/2023MNRAS.521.4056W} {521, 4056}

\bibitem[\protect\citeauthoryear{{Wu}, {McQuinn}, {Kannan}, {D'Aloisio}, {Bird}, {Marinacci}, {Dav{\'e}}  \& {Hernquist}}{{Wu} et~al.}{2019}]{wu19}
{Wu} X.,  {McQuinn} M.,  {Kannan} R.,  {D'Aloisio} A.,  {Bird} S.,  {Marinacci} F.,  {Dav{\'e}} R.,   {Hernquist} L.,  2019, \mn@doi [\mnras] {10.1093/mnras/stz2807}, \href {https://ui.adsabs.harvard.edu/abs/2019MNRAS.490.3177W} {490, 3177}

\bibitem[\protect\citeauthoryear{{Yang} et~al.,}{{Yang} et~al.}{2020}]{yang20}
{Yang} J.,  et~al., 2020, \mn@doi [\apjl] {10.3847/2041-8213/ab9c26}, \href {https://ui.adsabs.harvard.edu/abs/2020ApJ...897L..14Y} {897, L14}

\bibitem[\protect\citeauthoryear{{Zel'dovich}}{{Zel'dovich}}{1970}]{zeldovich70}
{Zel'dovich} Y.~B.,  1970, \aap, \href {https://ui.adsabs.harvard.edu/abs/1970A&A.....5...84Z} {5, 84}

\bibitem[\protect\citeauthoryear{{Zhu} et~al.,}{{Zhu} et~al.}{2022}]{zhu22}
{Zhu} Y.,  et~al., 2022, \mn@doi [\apj] {10.3847/1538-4357/ac6e60}, \href {https://ui.adsabs.harvard.edu/abs/2022ApJ...932...76Z} {932, 76}

\bibitem[\protect\citeauthoryear{{Zhu} et~al.,}{{Zhu} et~al.}{2023}]{zhu23}
{Zhu} Y.,  et~al., 2023, \mn@doi [arXiv e-prints] {10.48550/arXiv.2308.04614}, \href {https://ui.adsabs.harvard.edu/abs/2023arXiv230804614Z} {p. arXiv:2308.04614}

\makeatother
\end{thebibliography}






\bsp	
\label{lastpage}
\end{document}